\documentclass[preprint,nonatbib]{cuparticle}

\usepackage{apacite}
\usepackage{lineno,hyperref}
\usepackage{caption}
\usepackage{subcaption}
\usepackage{array}
\usepackage{multirow}
\usepackage{amssymb}
\usepackage{extarrows}
\usepackage{hyperref}
\usepackage{mathdots}
\usepackage[normalem]{ulem}
\usepackage{color}
\usepackage{media9}
\usepackage{placeins}
\usepackage{setspace}
\usepackage{graphicx}
\usepackage{calc}

\usepackage{enumitem}
\graphicspath{ {figures/} }

\usepackage{xpatch}
\makeatletter
\xpatchcmd\titlepage{\setcounter{page}\@ne}{}{}{}
\xpatchcmd\endtitlepage{\setcounter{page}\@ne}{}{}{}
\makeatother

\modulolinenumbers[5]
\setlist{leftmargin=*,rightmargin=\leftmargin}

\begin{document}
	\doublespacing
	\begin{titlepage}
		
		\authorheadline{}
		\runningtitle{}
		
		\title{Inspiration Hunter for Picture Design via Cut-transform-paste}
		\author{Gulce Bal Bozkurt}
		\author{Sibel Tari\corref{mycorrespondingauthor}}
		\address{Middle East Technical University, Ankara \ead{stari@metu.edu.tr}}
		
		\begin{abstract}
			{The digital era transformed design paradigms, blurring borders among exploration, ideation, and construction. Crowd generated big digital data emerged as a new material.
				There appears to be a need for new content creation tools which do not ignore complex interaction among traditionally separate stages and roles. 
				To address this need in picture design, we present an open-source platform called {\sl Inspiration Hunter}. It facilitates ideation via search-edit by supplying infinitely many dynamic possibilities available through the web without leaving the content creation environment, hence, seamlessly integrates exploration, ideation, and construction. User experiences show that the platform successfully enables reuse of web data both as a conceptual stimulant and as a material available for cut-transform-paste. Extending grammar-based approach to shape design, Inspiration Hunter mediates data-enabled collaborative creativity  beyond common intention, time or space.}
		\end{abstract}
		
		
		\maketitle
		
	\end{titlepage}	
	%%%%%%	
	\section{Introduction}
	\noindent 
	Today, modern design involves activities  that were not traditionally considered as part of designing.
	Several works promote  ``constructing/making" as  an integral part of the design process.
	The digital technology in particular seems to alter  paradigms, practices, materials and roles.
	Digital data seems to emerge as a new material.
	The concept of design as purposeful and goal-oriented manipulation of symbolic representations is giving way to interaction, imitation, collaboration in an endless iteration.
	
	In this emerging setting, Inspiration Hunter is a contemporary platform for picture design, extending the seminal work of shape grammars \cite{StinyBook}.
	In particular, it extends the capability of a content creation tool by providing services that link the content creation tool to a web-based search engine. 
	Its users can integrate retrieved images to their design via cut-transform-paste without having to leave their content creation environment. In addition to pose and color changes, the transform operations also includes style transformations facilitated by the Inspiration Hunter using its style adaptation modules.
	We believe that seamless integration of the inspiration source to the content creation tool is important so that users and designers can represent ideas visually with ease.
	In their extensive investigation on the effect of stimuli representation on ideation,  \citeA{sarkar_chakrabarti_AIEDAM} observe  that designers represent ideas predominantly nonverbally when triggered nonverbally, and verbally when triggered verbally. Hence, supporting the importance of pictorial trigger in pictorial design.

\begin{figure}[h]
		\centering
		\begin{tabular}{c}
			\includegraphics[width=.15\textwidth]{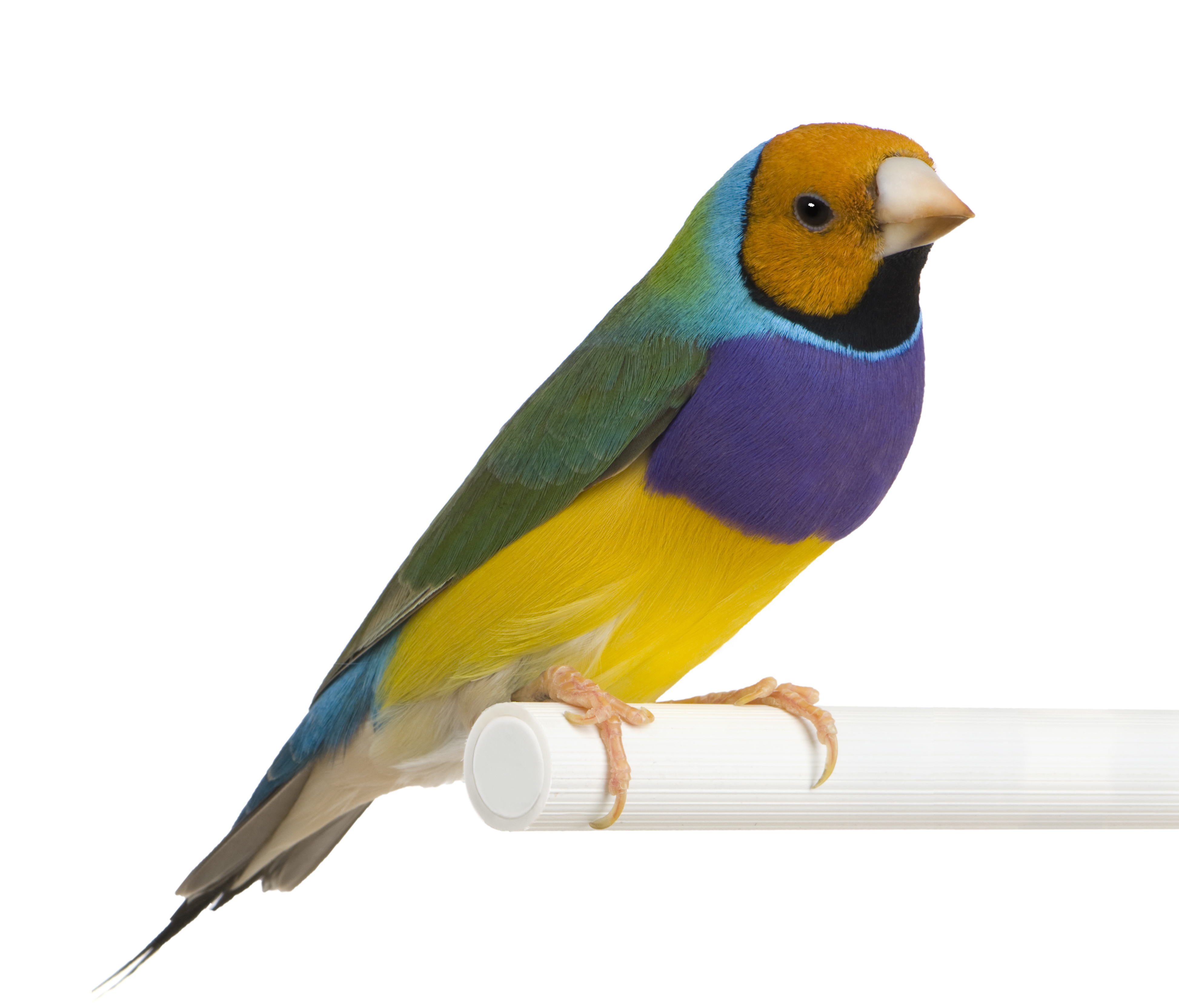} \\
			(a) \\
			\includegraphics[width=.6\textwidth]{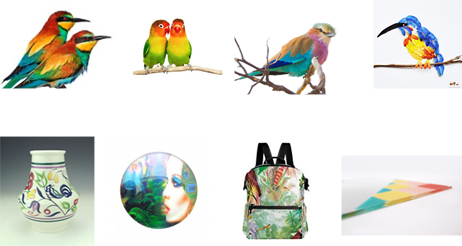} \\
			(b) \\
		\end{tabular}
		\caption{ Given a bird image as a query (a), both relevant and irrelevant results might pop up (b); we consider this surprise factor a  blissful source for divergence}
		\label{fig:divergencesource_1}
	\end{figure}

	As the web search engine (the inspiration source), we chose Google Images \cite{GoogleImages} because it is the most widespread user interface to search visual references from the web. Since 2010, the interface allows search via  image in addition to search via keywords. Over 10 billion indexed images feed the Google search engine. Hence, designers can explore the practically limitless source of Google Images using their current design or scribbles, any image from Google, an image patch or some text specifying a concept. Google Reverse Image Search Engine is not transparent; any result might pop up. For example, given an image of a bird as a query, the engine returns not only similar birds  but also completely irrelevant objects sharing the texture or color distribution of the query image (Figure~\ref{fig:divergencesource_1}). 
	
	We see this  as a blessing than a blight. This is because  uninteresting  ideas may contribute to the generation of interesting ones.  Among others, several recent studies \cite{Sawyer,Harrison,Cagan2019} reviewed in the Literature section support this view.
	Therefore, while taking advantage of massive amount of  content, we deliberately avoid data analytics and filtration criteria. Note that in contemporary work on shape grammars, the filtration is also achieved as a result of iterative seeing and doing, where  possible parts/shapes are captured either parametrically, e.g., \citeA{ensari_ozkar_2018,Grasl2013,Keles2010,prats_earl_garner_jowers_2006} or via implicit codings \citeA{Keles2012}. In contrast, the possibilities in our picture design framework is chance-based samplings from practically limitless ever-changing dynamic source.	
	
	The rest of the paper is organized as follows:  In \S \ref{sec:lit} we discuss  supporting work from recent literature. 
	In \S \ref{sec:overview}, we describe general features of the system; implementation details are given in \S \ref{sec:tech}. 
	Copyright and crediting issues are discussed in \S \ref{sec:copyright} and 
	user experiences along with sample results are demonstrated in \S \ref{sec:exp}.   
	
	A demo of the Inspiration Hunter is provided as a supplementary $mp4$ file $inspvideo.mp4$ accompanying this paper.	
	
	The complete software will be made publicly available as a part of this  article.  
	
	\section{Literature}
	\label{sec:lit}
	\noindent 
	Contemporary studies speak of creative design as divergent and unpredictable process intertwined with the physical act of constructing.
	For example, in an interesting recent empirical study based on an interview with MFA students in painting programs complemented with studio observations,  \citeA{Sawyer} reports that the creative process seems to be wandering, unpredictable, nonlinear and embedded in the process of constructing. 
	A further  interesting observation reported by \citeA{Sawyer} is  that his empirical study fails to provide  any evidence that could suggest that an attempt to be original may play a role in the creative processes of the subjects. In a similar vein, in the context of shape grammars based computational design, \citeA{Harrison} report that unsuccessful modifications were often more useful for positively identifying essential relationships than successful ones.
	\citeA{Cagan2019} explore whether inspirational stimuli can come from unlikely resources such as crowd generated data.  
	Furthermore, again in the context of shape and making grammars, researchers in the last decade promote constructing as  an integral part of the design process; addressing the complex interplay between design and construction, e.g. \citeA{KnightVardouli, GursoyOzkar, ElZanfaly}.
	
	Lastly, as we mentioned in the Introduction, 
	\citeA{sarkar_chakrabarti_AIEDAM} extensively  investigate  the effect of stimuli representation on ideation. They compare six different representations including image  and video. Their results not only  indicate that the image is the most effective representation outperforming even the video but also highlights the importance of nonverbal trigger if the designers are to represent ideas nonverbally.

	\section{System Overview}
	\label{sec:overview}
	\noindent  
	The default content creation environment for  Inspiration Hunter is Adobe Photoshop or Illustrator, which are widely used by both amateurs and professionals. 
	In 2011, Adobe Systems released a service called Creative Cloud which  allows users to access  application software in the Creative Suite.
	Nevertheless, following the implementation details provided in \S \ref{sec:tech}, modifications can be made to adapt the platform to another content creation tool that  allows  users to access  application software to extend its functionality. For example, GIMP - GNU Image Manipulation Program. 
	
	The system has two layers: server and client. The main functionality resides in the server layer which is stored in a remote server independent of the  personal computer of the designers (will be referred as platform users).
	Without delving into technical details, it is important to note that we used the so called {\sl client-server architecture}. This is mainly  because we  want  platform users to be blind to our development platform, resource requirements or  future updates. They are able to use  Inspiration Hunter in their own personal computers without being obliged to have or prepare a special platform.  For example, they are able to use a service that we have developed using Java without needing Java in their personal computers. Likewise, computationally loaded services  such as style transfer 
	can run on their personal computer even if their computers do not  have Graphical Processing Units.  
	Yet another motivation for client-server architecture is reusability: With client-server separation, our interfaces can also be used by systems other than Inspiration Hunter.

	A platform user's PC should contain our  custom-built extension to Adobe Creative Cloud. 
	These include 
	image download,  image-based hunt, keyword-based hunt, transferring the style of a selected image to the active image document, and applying known artistic styles to the current design. 

	Our extensions make it possible for  platform users to search the web  without leaving their design environment. Each time, they  may pick a result and reload it as a query. They may cut a piece, maybe a color patch, and reload it as a query instead of describing a color with its name. During the process, a new concept might pop up possibly motivating a user to enrich the search with a keyword. Examples are discussed in \S \ref{sec:exp}.
	
	Platform users do not need to upload or save their current design. The system handles this process in the background. By clicking an image among the retrieved search results, users can download a selected image which is then automatically opened as a new document.  
	We emphasize that whenever an image is downloaded, the system automatically removes all the associated metadata and renames the file using the system timestamp. This is important because when a downloaded image is used for a new hunt, the retrievals will be purely visual-based unless users themselves purposely enrich the  hunt with textual keywords.
	
\begin{figure}[h]
		\centering
		\begin{tabular}{cc}
			\includegraphics[width=0.47\textwidth]{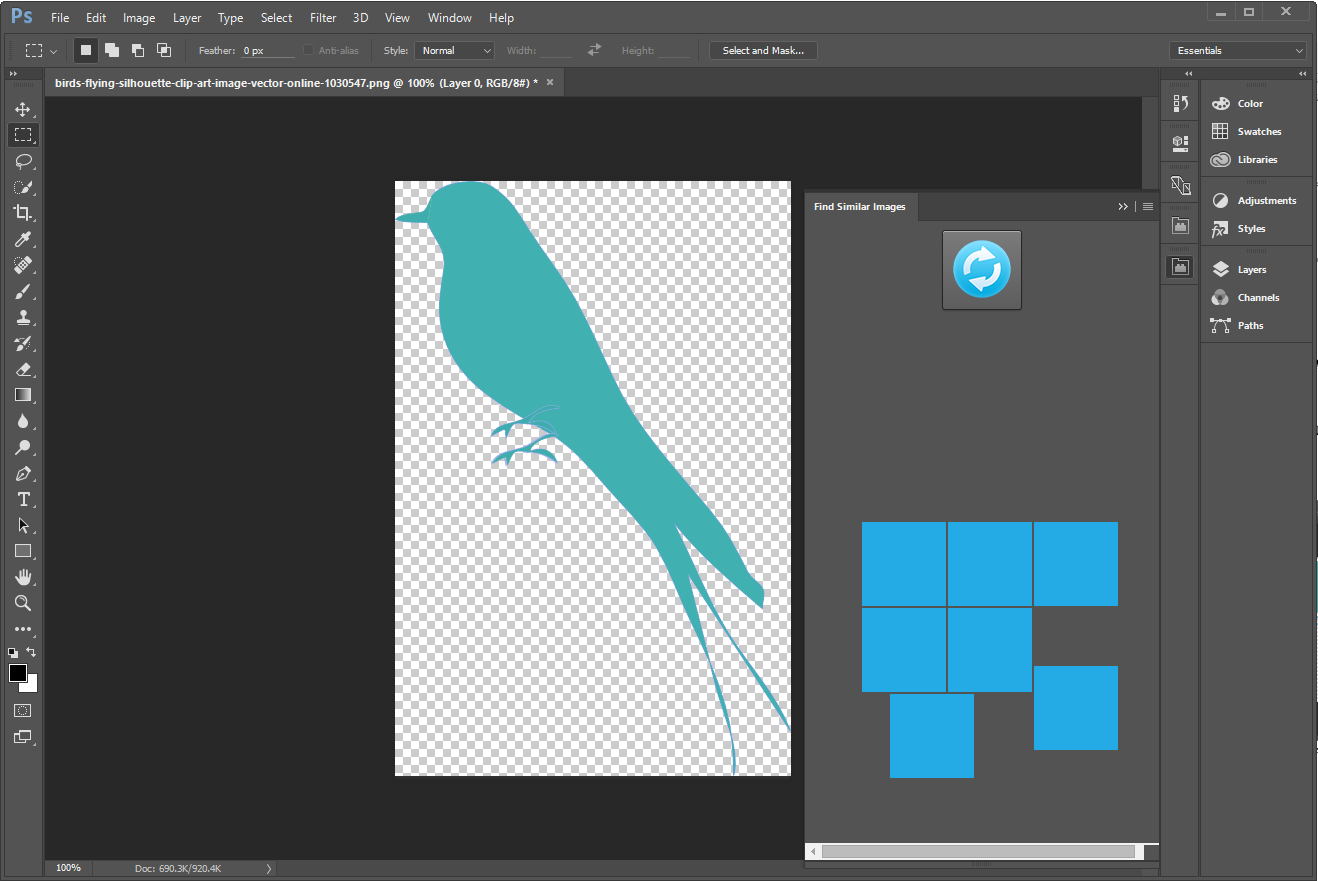} &
			\includegraphics[width=0.47\textwidth]{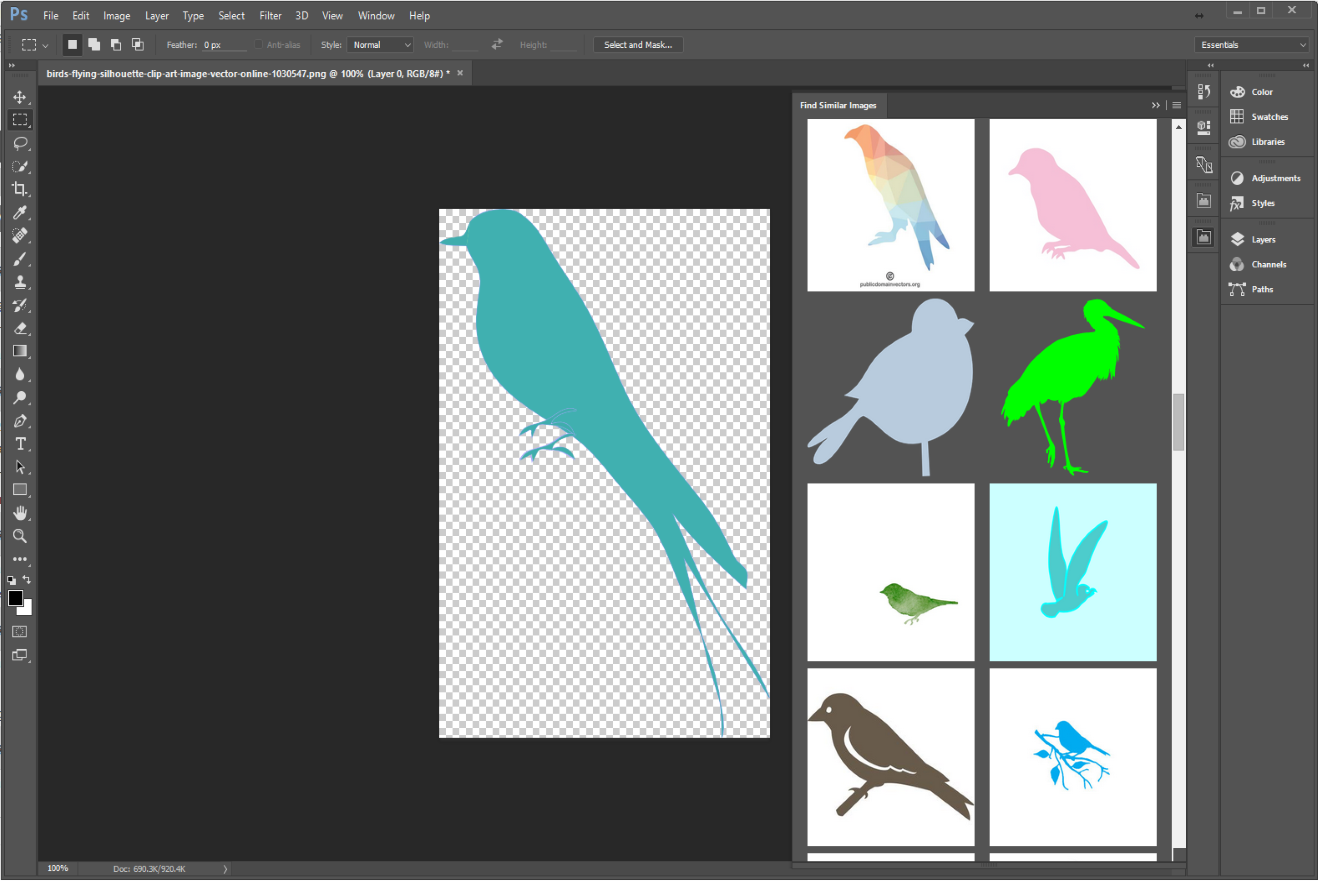}\\
			(a) & (b) 
		\end{tabular}
		\caption{Image-based search interface}
		\label{fig:searchInterfaces}
	\end{figure}

	The image-based hunt interface is illustrated  in Figure~\ref{fig:searchInterfaces}. Just by clicking the blue refresh button in Figure~\ref{fig:searchInterfaces} (a) and the system retrieves similar images and displays them as shown in  Figure~\ref{fig:searchInterfaces} (b).  Style adaptation is achieved using Neural Style Transfer (\S \ref{app:imagestyle}).  
	We strongly suggest readers to view accompanying supplementary video  $inspvideo.mp4$. 

	\section{Crediting and Copyright}
	\label{sec:copyright}
	\noindent Under fair use law, it is acceptable to reuse some material if the purpose  is for education, scholarship or research. 
	Transformative use is a later addition to fair use law. 
	A work is considered transformative if it uses the copyrighted material  in completely new or unexpected ways. 
	Non-profit purposes further favor fair use.  
	Though it is platform users' responsibility to adhere to the terms of current copyright laws, 
	Inspiration Hunter provides two services
	to reduce their workload (whether the purpose is non-profit or for-profit).
	
	Firstly, the web page address for the downloaded image and access time information is displayed as a text (See \S \ref{sec:appendix_visuals}). Depending on the nature of their use, platform users can copy this information to give proper credit/citation  or to contact for permission. Secondly, during the search platform users can select among several  label options such as {\sl reuse with modification, reuse, noncommercial reuse with modification, noncommercial reuse}, and {\sl not filtered by license}. The default for Google search is to bring all types of photos with all types of usage rights. However, it may be important to eliminate some photos and use filters to organize retrieval results based on reuse purpose. This is especially useful if the product is intended as for-profit and the platform user prefers to avoid copyright violation risk.

	\section{Platform User Experience Test and Sample Results}
	\label{sec:exp}
	\noindent We tested Inspiration Hunter with the help of volunteer platform users.
	All users are instructed to produce their content by forming collages via cut-transform-paste from the content offered by the system rather than generating from scratch. 
	
	\subsection{Tests Using Style Adaptation Tool}
\begin{figure}[h]
		\centering
		\begin{tabular}[t]{cccc}
			\includegraphics[trim=0 0 0 0, clip,height=0.155\linewidth]{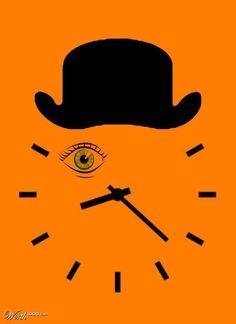} &  $\cdots$ & $\longleftarrow$ 	 
			\includegraphics[trim=150 220 40 30, clip,height=0.155\textwidth]{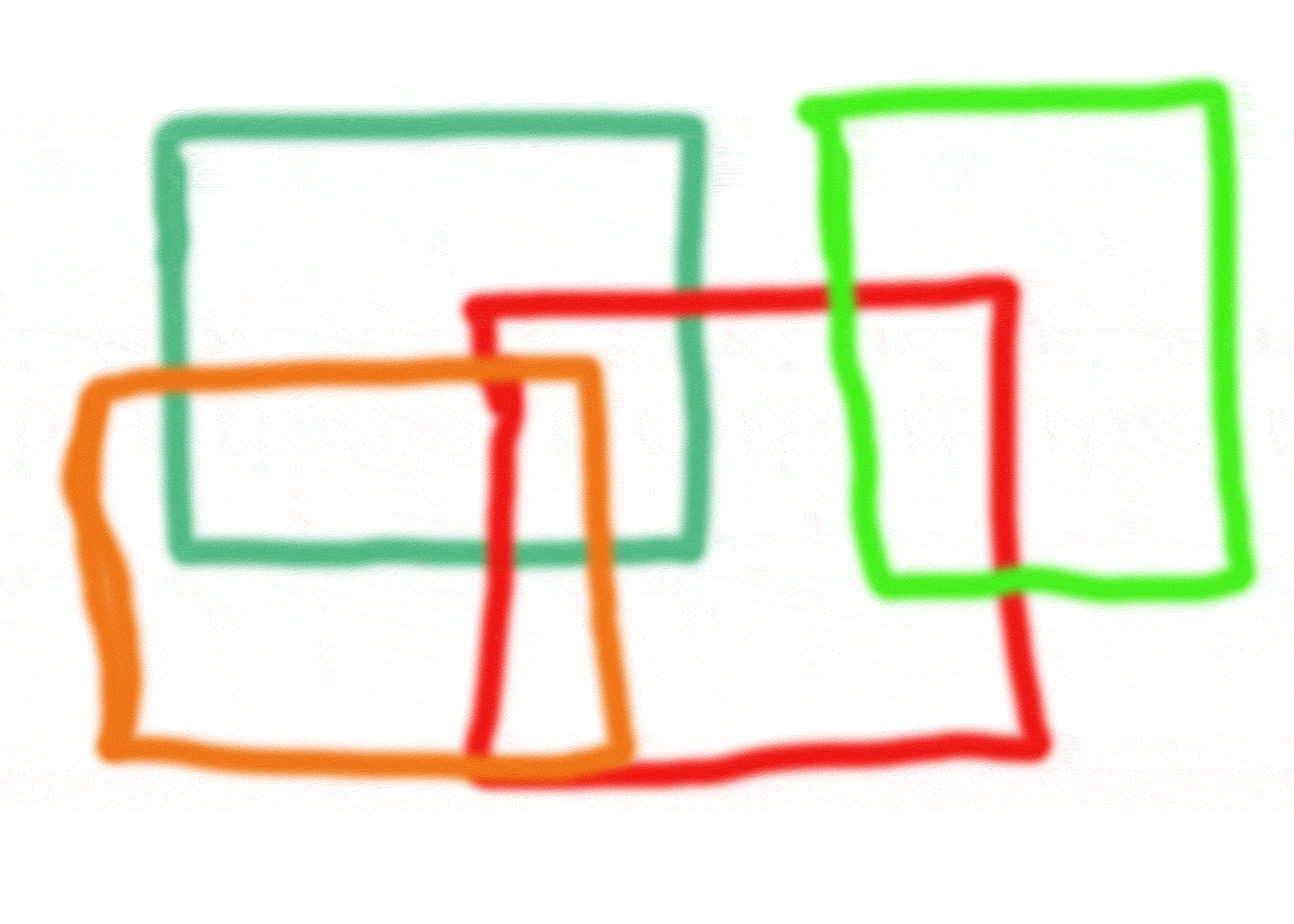} \\ \\ 
		\end{tabular}
		\begin{tabular}[t]{ccc}
			\includegraphics[trim=0 0 0 0, clip,height=0.53\textwidth]{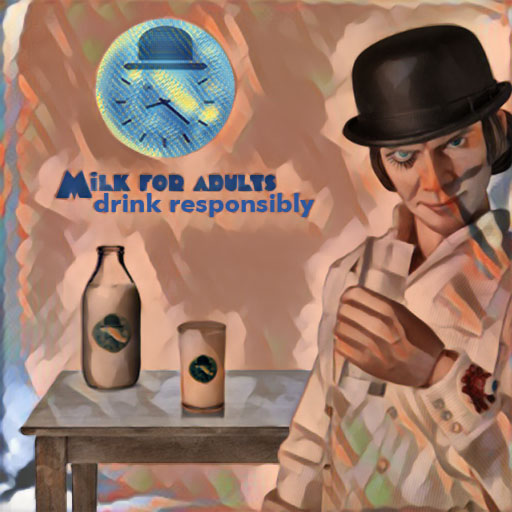} & &
			\includegraphics[trim=0 0 0 0, clip,height=0.53\textwidth]{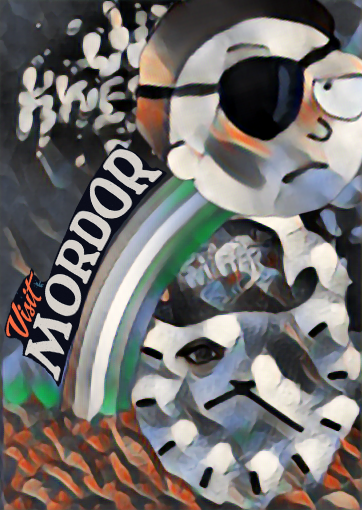}  	
		\end{tabular}
		\caption{Two cut-transform-paste designs (bottom row) diverging from the same starting point (top row)}
		\label{fig:products_4}
	\end{figure}

	\noindent 
	We provided users with a {\sl Clockwork Orange} logo (the top left in Figure~\ref{fig:products_4}) and suggested them to reuse this logo in their final product. Users  are encouraged to use the style adaptation tool. 
	Final products are shown in the second row of Figure~\ref{fig:products_4}. 
	The first user who designed a poster for a milk advertisement had watched the movie {\sl A Clockwork Orange} whereas the other user had not.
	We note that this logo was also retrieved by Inspiration Hunter as a response to a sequence of image hunts starting with the scribbled squares by the authors (the top right in Figure~\ref{fig:products_4}). Design process details are given  in \S~\ref{ssec:Exp3a} and \S~\ref{ssec:Exp3b}.

	\begin{figure}[h]
		\centering
		\begin{tabular}[t]{cccccc}
			\includegraphics[trim= 0 0 0 0, clip, height=0.15\textwidth]{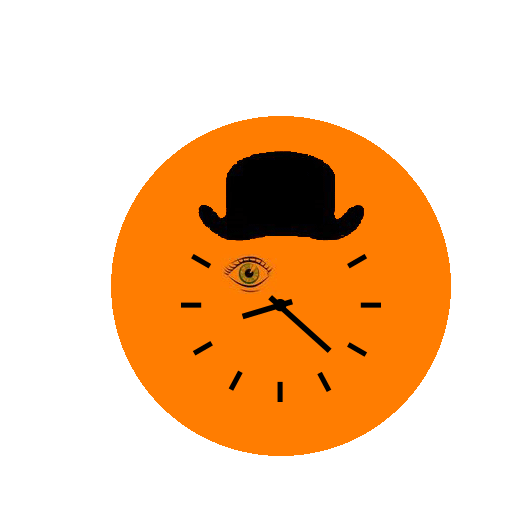} &
			\includegraphics[trim= 0 0 0 0 clip,height=0.15\textwidth]{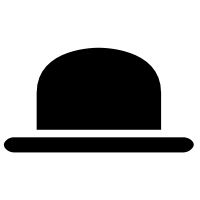} &
			\includegraphics[trim= 0 0 0 0, clip, height=0.15\textwidth]{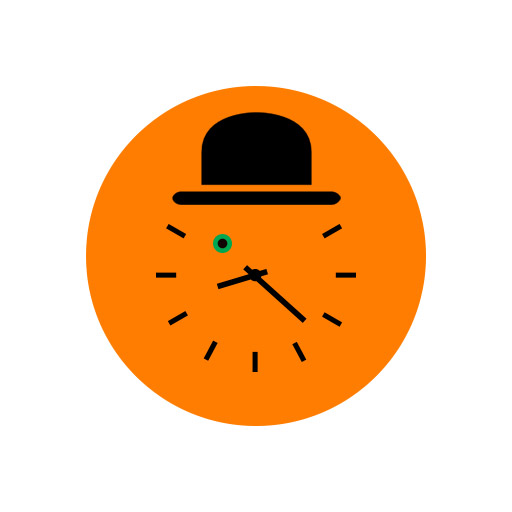} &
			\includegraphics[trim= 100 0 100 0, clip,height=0.15\textwidth]{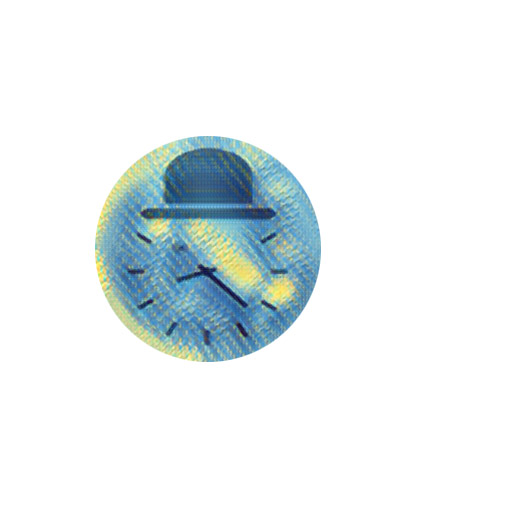}&
			\includegraphics[trim= 0 70 260 70, clip, height=0.15\textwidth]{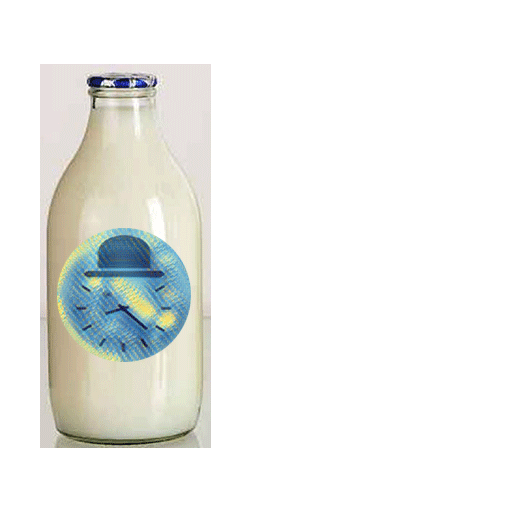}&
			\includegraphics[height=0.15\textwidth]{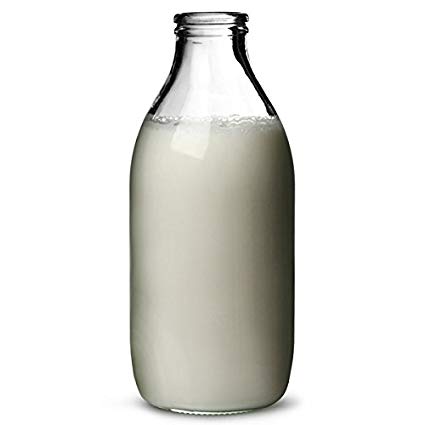} \\
			{\tiny 1} & {\tiny 2} & {\tiny 3} & {\tiny 5} & {\tiny 7}&  {\tiny 8} \\\\
		\end{tabular}
		\begin{tabular}[t]{ccccc}	
			\includegraphics[height=0.15\textwidth]{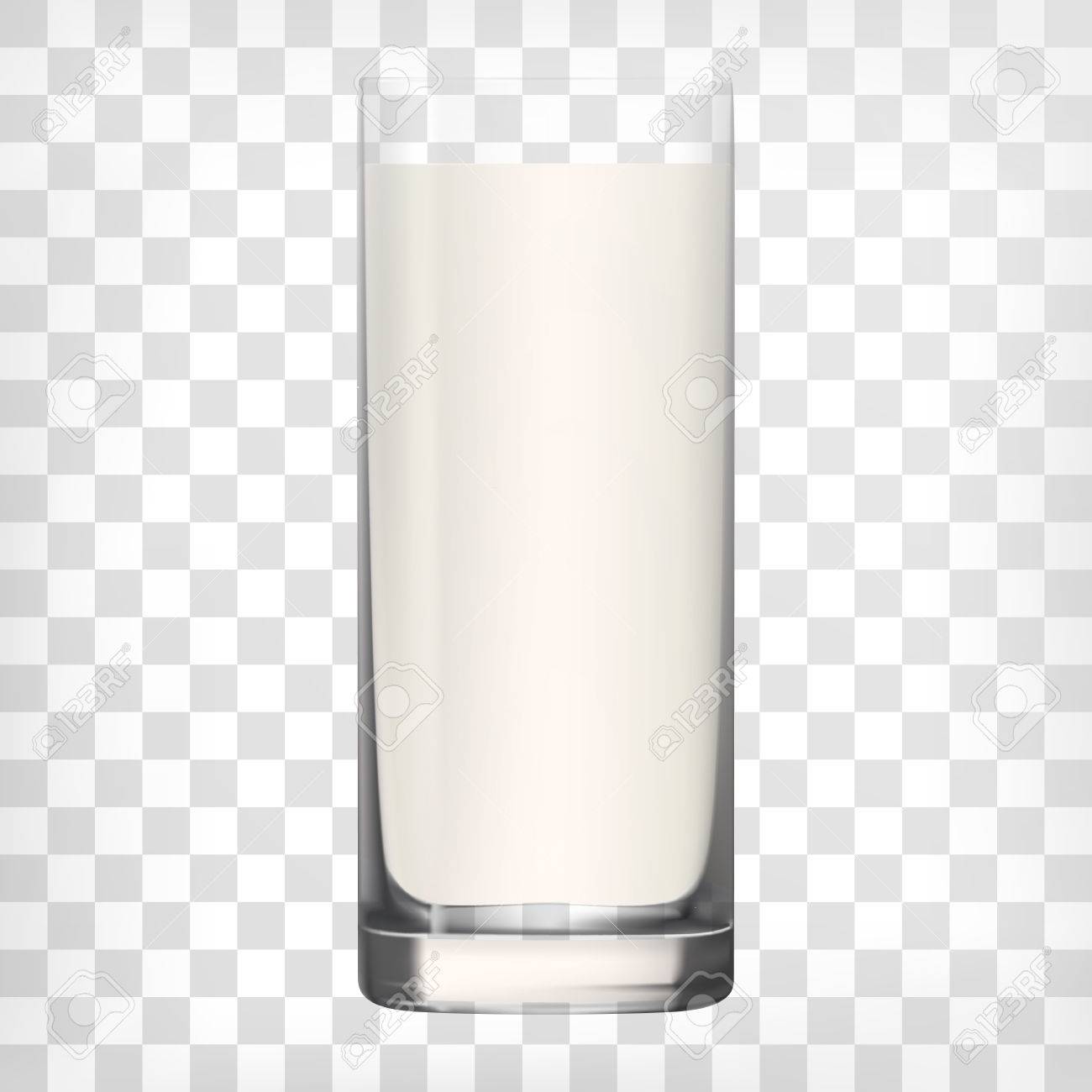}&
			\includegraphics[trim= 0 0 0 30, clip, height=0.15\textwidth]{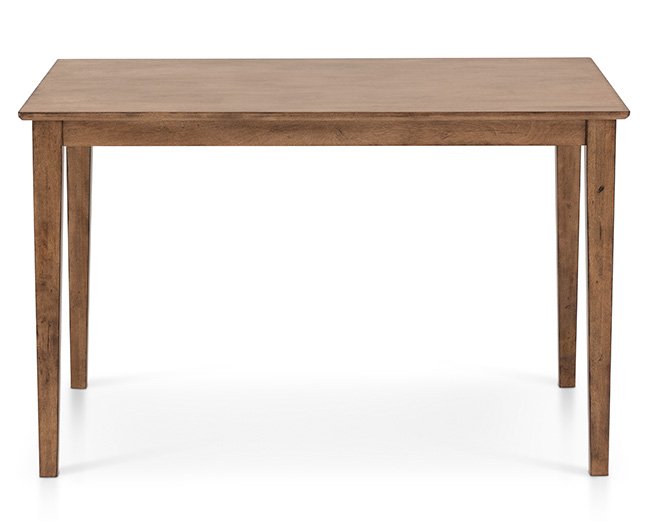}&
			\includegraphics[height=0.165\textwidth]{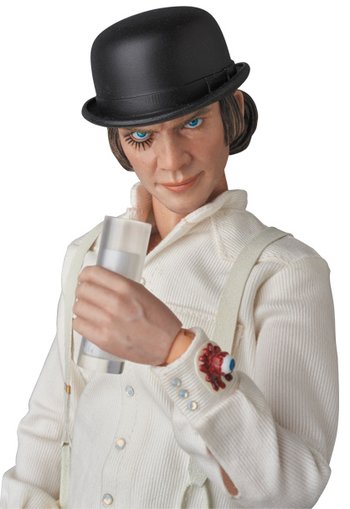}&
			\includegraphics[height=0.165\textwidth]{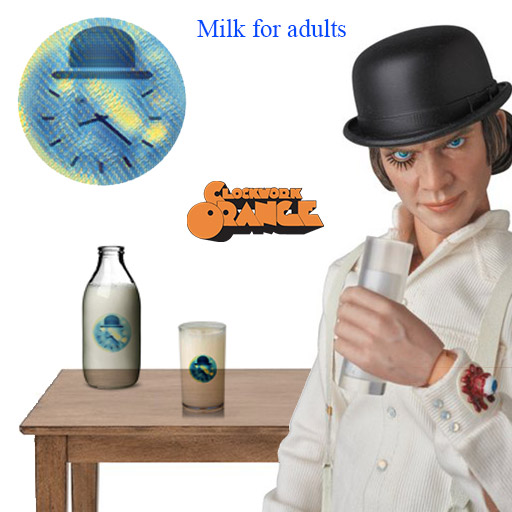} &
			\includegraphics[height=0.165\textwidth]{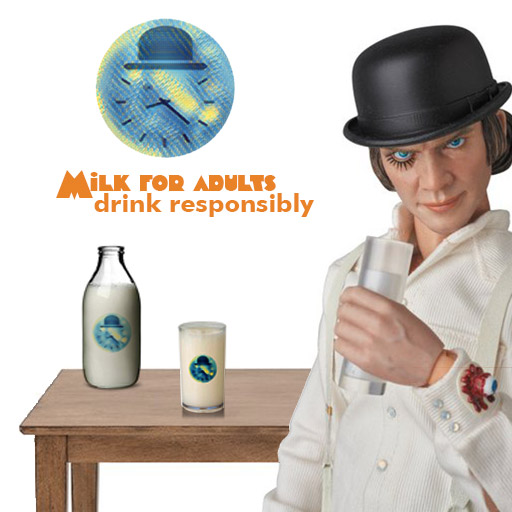}\\
			{\tiny 9}&{\tiny 10}& {\tiny 11}&{\tiny 13} &{\tiny 14}\\\\
		\end{tabular}
		\begin{tabular}[t]{cccccc}	
			\includegraphics[height=0.165\textwidth]{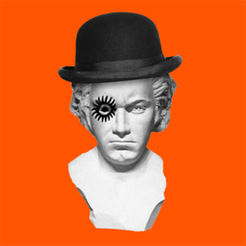}&
			\includegraphics[trim= 0 0 0 0, clip, height=0.165\textwidth]{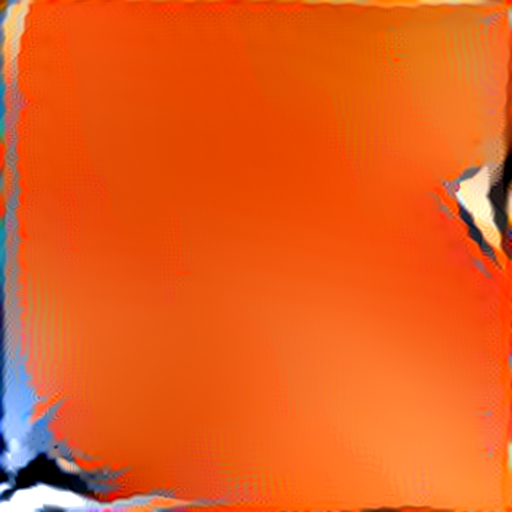} &
			\includegraphics[trim= 0 0 0 0, clip, height=0.165\textwidth]{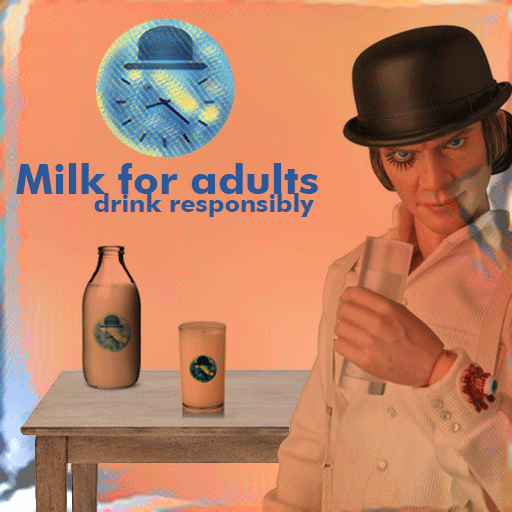} &
			\includegraphics[trim= 0 0 0 0, clip, height=0.165\textwidth]{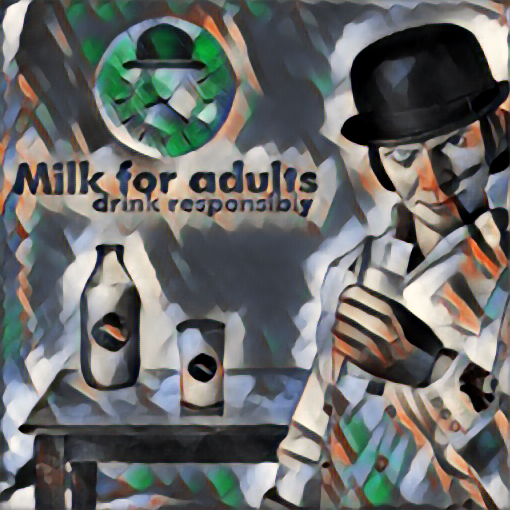} &
			\includegraphics[trim= 0 0 0 0, clip, height=0.165\textwidth]{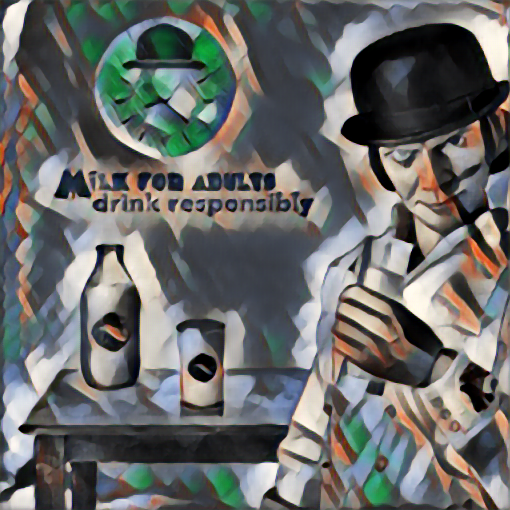} \\
			{\tiny 15}&{\tiny 16}& {\tiny 17}&{\tiny 18} &{\tiny 19}\\\\
		\end{tabular}
		\caption{A Clockwork Orange - I}
		\label{fig:experiment4A}
	\end{figure}

	\subsubsection{A Clockwork Orange - I}
	\label{ssec:Exp3a}
	\noindent  The evolution of the design is depicted in Figure~\ref{fig:experiment4A}.
	The user did a keyword-based search by using {\sl hat icon} as the keyword. In the meantime, she cut a circular section (1). 
	Among the retrievals to her keyword-based search, she picked the hat shown in (2) and replaced the original hat with the new one. 
	In the meantime, she also simplified the eye yielding  (3). After performing style transfer, she decided to design a coin and formed  (5) and used it for image-based search. The search results did not include any coin-like  images so she gave up the coin design idea. 
	Among the retrieval results, she noticed an image showing Alex holding a glass of milk (11). 
	
	This was a key moment: She decided to design a milk brand and use the coin-like form as the new brand's logo. 
	Hence, at the next step, she queried with the keyword {\sl milk in bottle}. She cut a bootle and pasted the logo on the bottle (7). She used the keyword-based search again to find a more aesthetic bootle.  She picked a new bottle (8).   She continued search using keywords {\sl milk in glass} and {\sl table} selecting (9) and (10). She used  keyword {\sl clockwork orange} and selected a text. She liked the font of the text, hence, downloaded it. She combined all  digital material on the collage, manually inserting a text {\sl Milk for adults} (13). She later revised the text content and its font using previously dowloaded fonts. She decided to change the background layer. She took the background of (15) as shown in (16) and applied it to the active collage. Then she changed the opacity yielding (17).  She applied style transfer (18) and changed the font (19). Finally, she decreased the opacity yielding the final product as shown in Figure~\ref{fig:products_4}.

	\subsubsection{A Clockwork Orange - II}
	\label{ssec:Exp3b}
\begin{figure}[h]
		\centering
		\begin{tabular}[t]{ccccccc}
			{\includegraphics[height=0.161\linewidth]{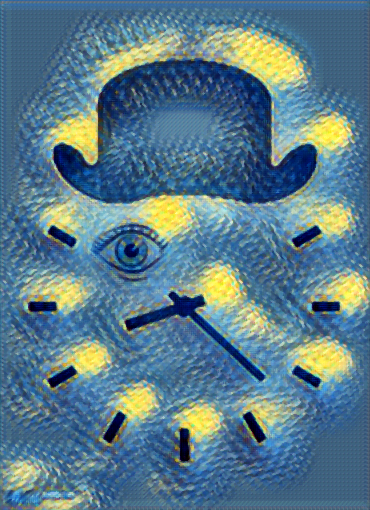}} &
			{\includegraphics[height=0.161\linewidth]{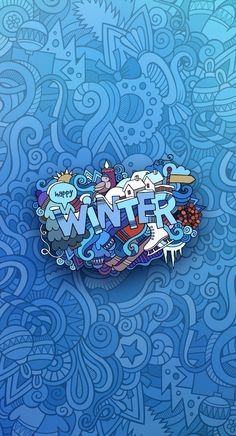}}&
			{\includegraphics[height=0.161\linewidth]{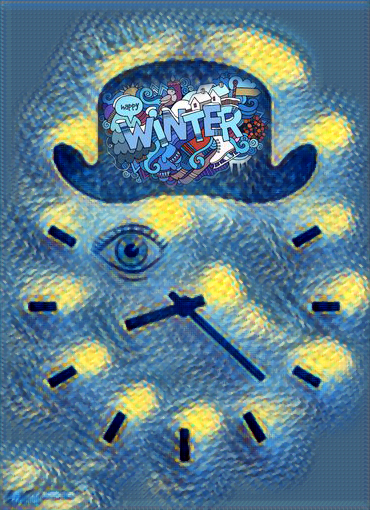}}&
			{\includegraphics[height=0.161\linewidth]{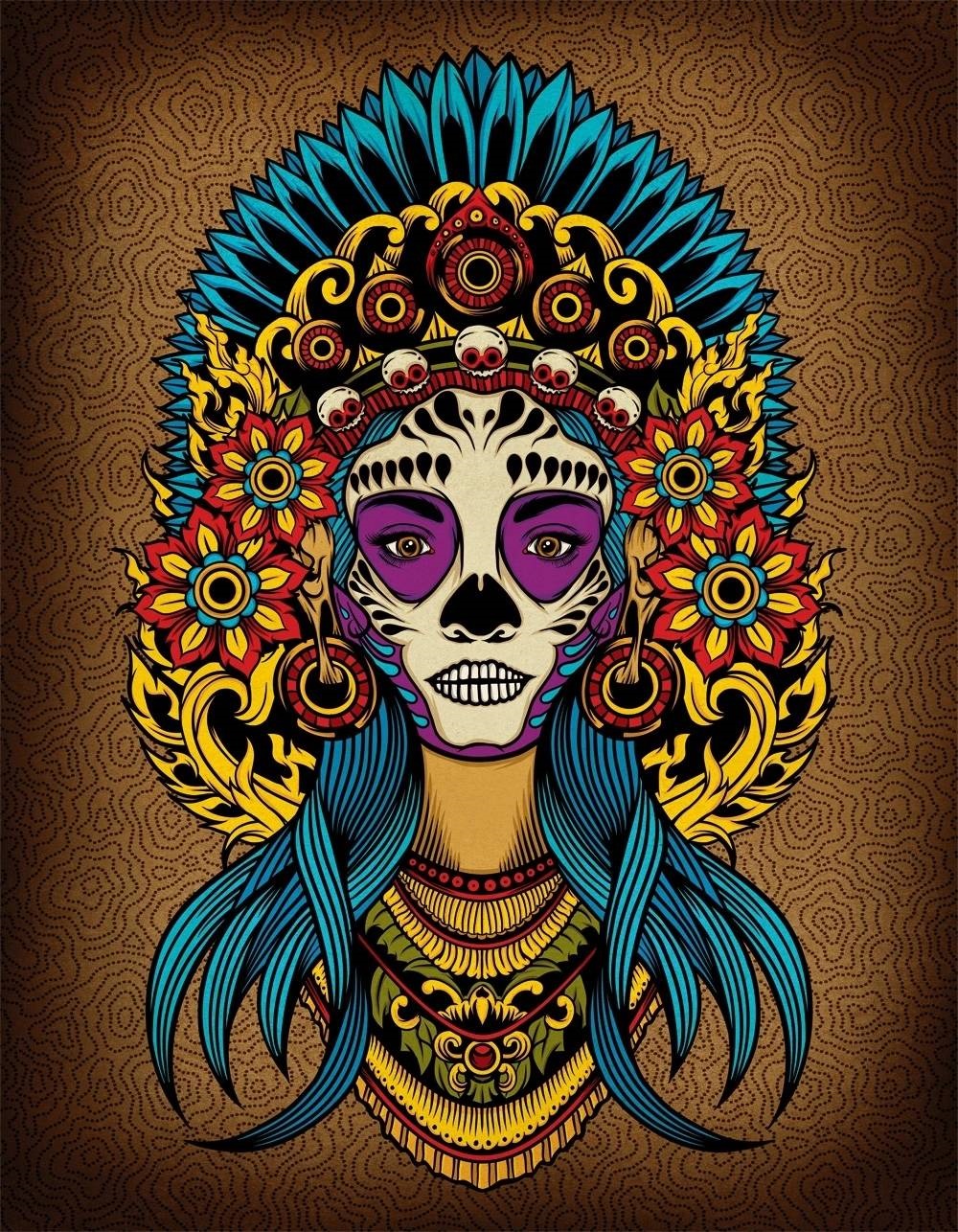}}&
			{\includegraphics[height=0.161\linewidth]{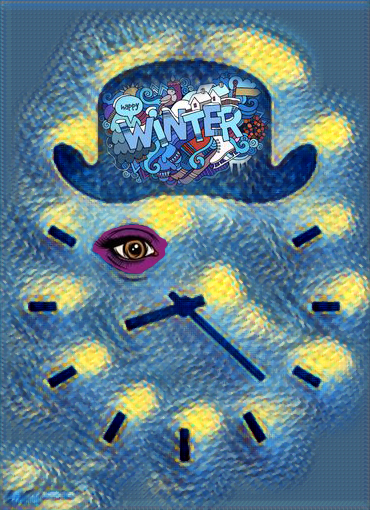}}&
			{\includegraphics[height=0.161\linewidth]{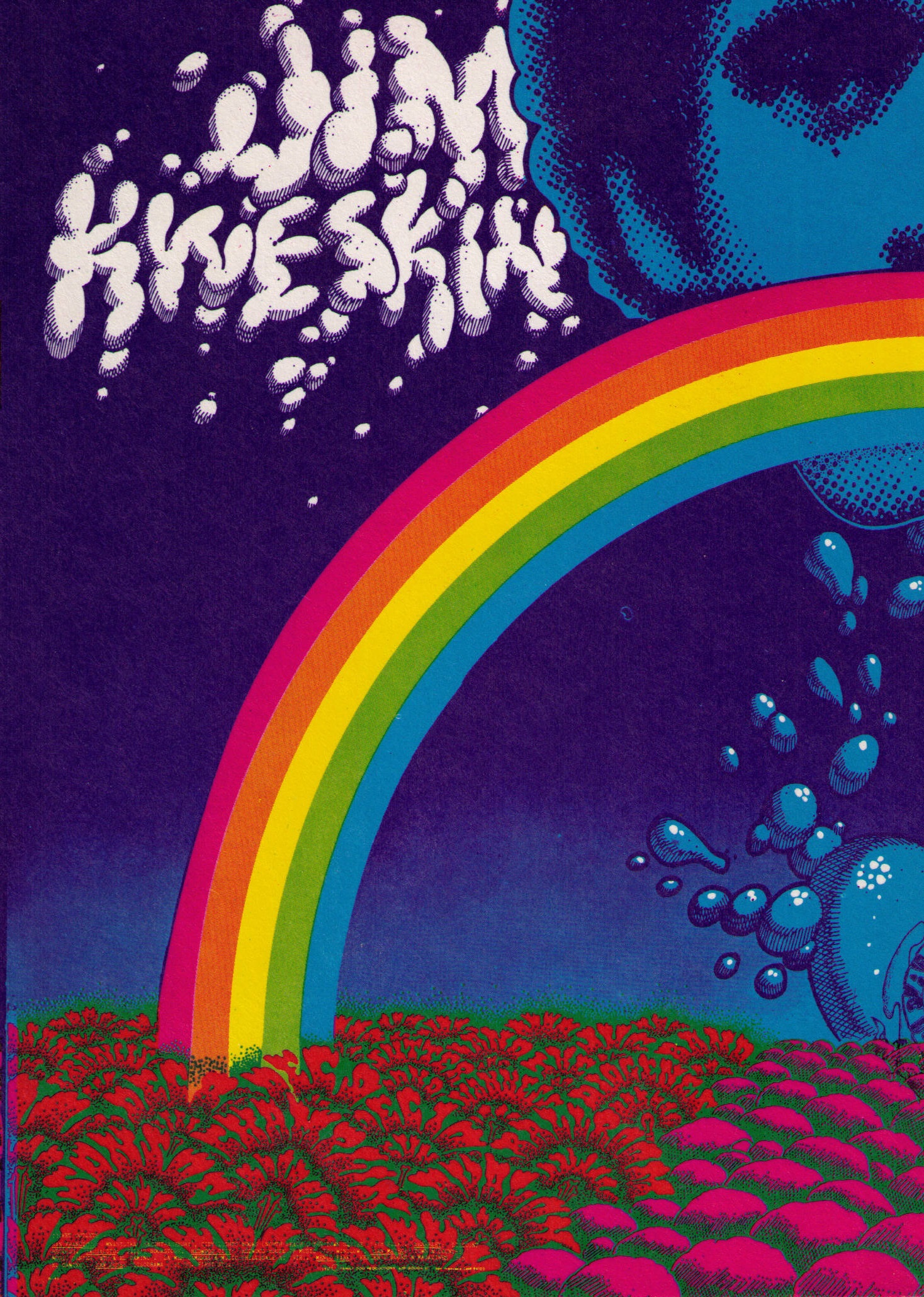}}&
			{\includegraphics[height=0.161\linewidth]{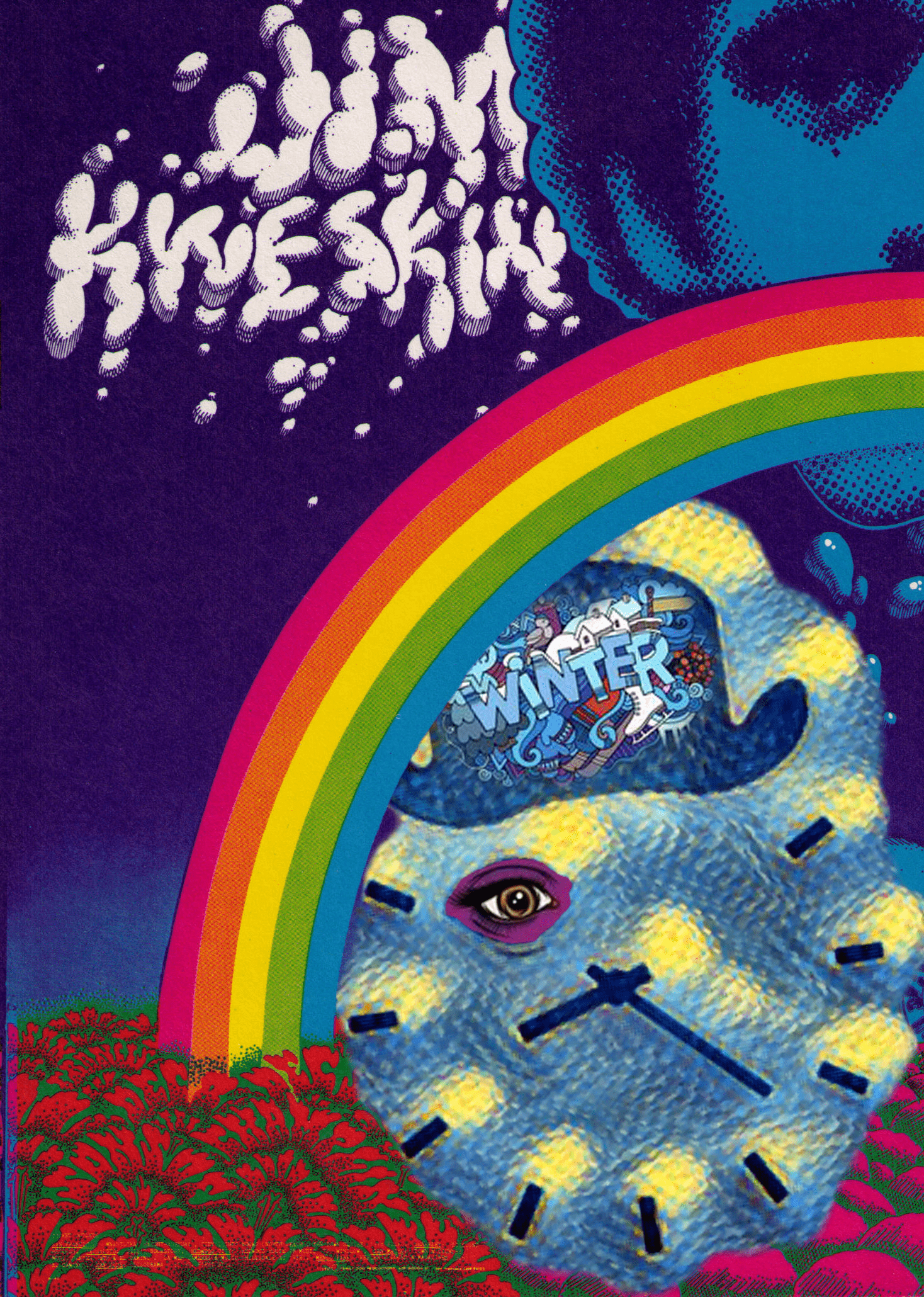}} \\
			{\tiny 1} & {\tiny 2} & {\tiny 3} & {\tiny 4} & {\tiny 5} & {\tiny 6} & {\tiny 7} \\
		\end{tabular} 			
		\begin{tabular}[t]{ccccccc}								
			{\includegraphics[height=0.161\linewidth]{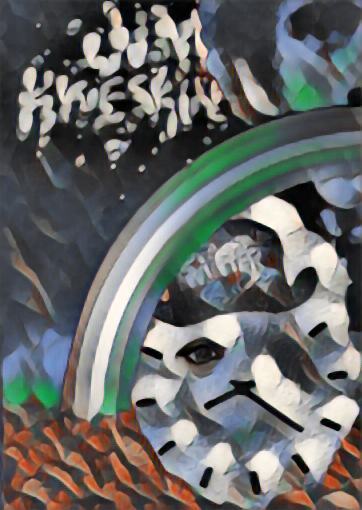}} &
			{\includegraphics[height=0.161\linewidth]{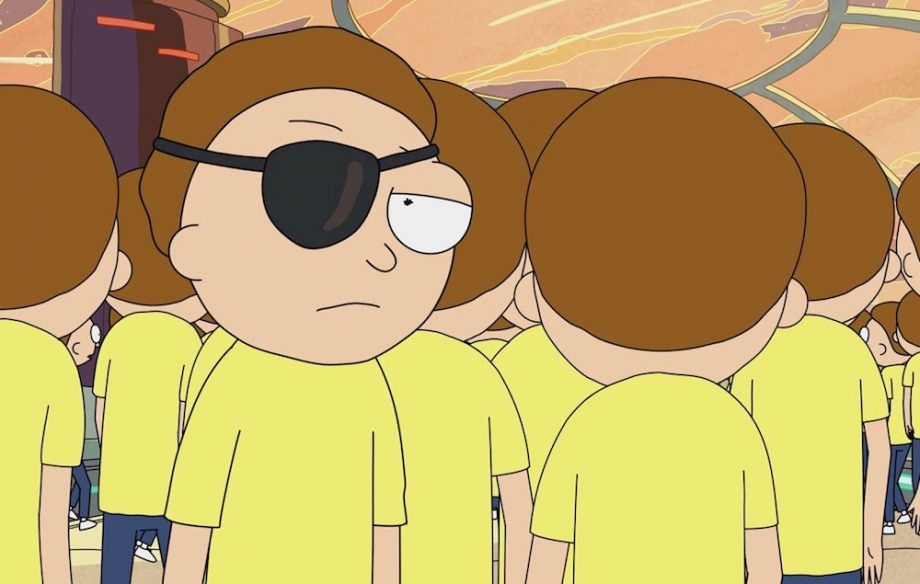}}&
			{\includegraphics[height=0.161\linewidth]{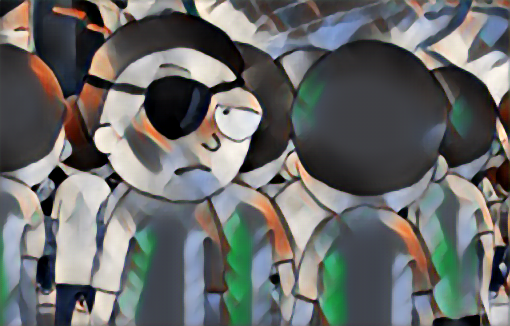}}&
			{\includegraphics[height=0.161\linewidth]{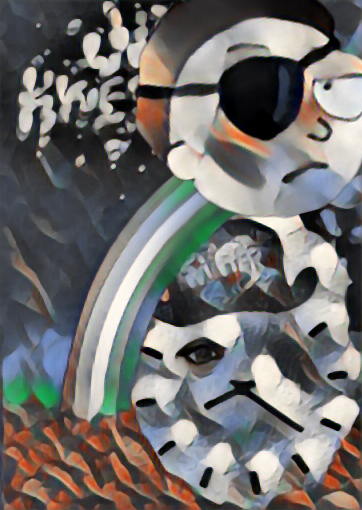}}&
			{\includegraphics[height=0.161\linewidth]{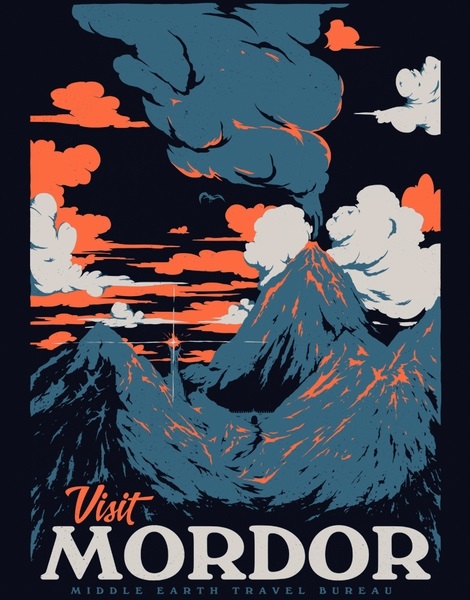}}\\
			%19& 20 & 21& 22&23 
			{\tiny 8} & {\tiny 9} & {\tiny 10} & {\tiny 11} & {\tiny 12} 
			%&&(b) \\
			% \hskip 1cm $\leftarrow$ \mbox{} 
		\end{tabular}
		\caption{A Clockwork Orange - II}
		\label{fig:experiment4B}
	\end{figure}

	\noindent The evolution of the design is depicted in Figure~\ref{fig:experiment4B}.  
	The user  started by applying a style change yielding (1). Using (1) as a probe retrieved (2).  The central content from (2) is cut and pasted resulting in (3). Using (3) as a query, the user retrieved  (4). Then he used the death goddess's purple eye to  replace the single eye in the active collage yielding (5).  Search with (5) retrieved (6), a Jim Kweskin poster. He replaced evolved clockwork orange from the active collage to the bottom right corner of the Kweskin poster where there is a suitable space under the rainbow yielding the new active collage shown in (7). After applying style change as shown in (8), the evolved clockwork orange reminded him the Rick character from {\sl Rick and Morty}. 
	At this stage, he   used keyword-based hunt using   {\sl Rick} and {\sl Morty} as keywords, which retrieved (9). He transferred the style of the the retrieved image using the style of the active collage yielding (10). He cut the one-eyed round face (Morty) and pasted it on the active collage (11). Using the new collage as a query retrieved (12), a {\sl Visit Mordor} poster. Finally, he cut the text {\sl Visit Mordor} and pasted it in a curved form along the rainbow, yielding the final product as shown in Figure~\ref{fig:products_4}.  
	
\begin{figure}[h]
		\centering
		\begin{tabular}{c}
			\includegraphics[width=0.86\textwidth]{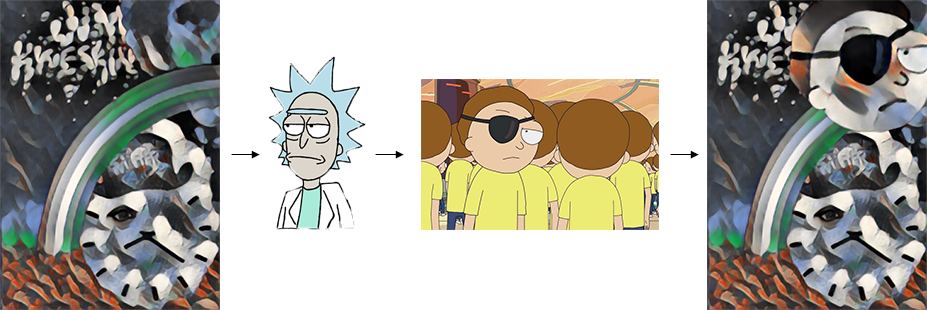} \\
		\end{tabular}
		\caption{Divergence and convergence through user's perception}
		\smallskip
		\label{fig:divergencesource_rick}
	\end{figure}

	This user experience exhibits a quite interesting source of divergence and convergence  with the evolved clockwork orange reminding the user completely unrelated cartoon character (Figure~\ref{fig:divergencesource_rick}). 

	\subsection{Tests without Using Style Adaptation Tool}
\begin{figure}[htb]
		\centering
		\begin{tabular}[t]{c} 
			\includegraphics[trim=0 0 0 0, clip,width=0.95\textwidth]{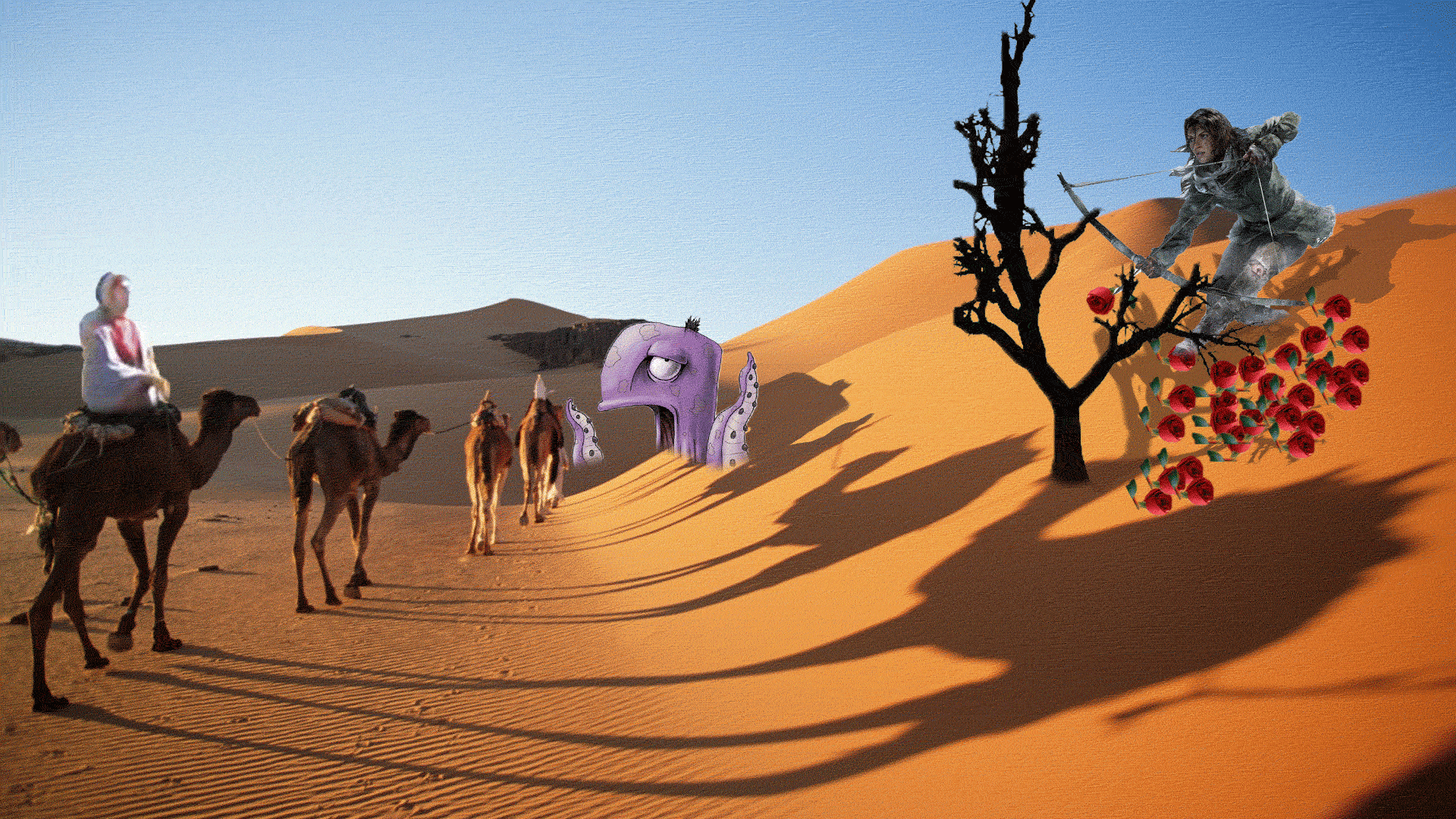} 
			% 	 \includegraphics[trim=0 0 0 0, clip,width=0.7\textwidth]{figures/dolphin_25} 
			%Lara Croft; desert & butterfly
		\end{tabular}
		\caption{Desert Monster: A cut-transform-paste design produced using Inspiration Hunter without style adaptation tool}
		\label{fig:product_lara}
	\end{figure}

\begin{figure}[htb]
		\centering
		\begin{tabular}[t]{c} 
			\includegraphics[trim=0 0 0 0, clip,width=0.8\textwidth]{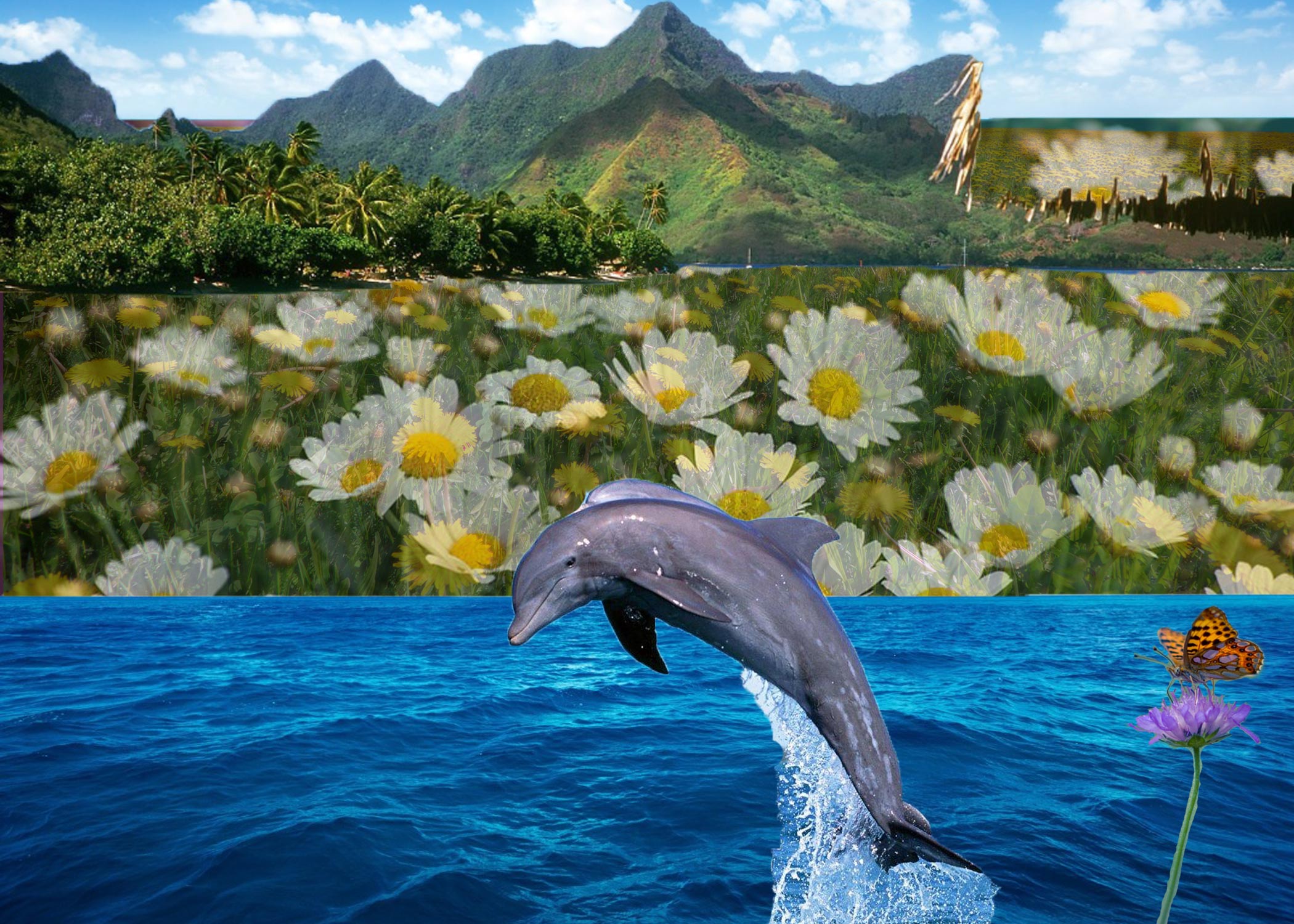} 
			%Lara Croft; desert & butterfly
		\end{tabular}
		\caption{Butterfly: A cut-transform-paste design produced using Inspiration Hunter without style adaptation tool}
		\label{fig:product_butterfly}
	\end{figure}

	\noindent In addition to turning off style adaptation tool, users are asked to start with a keyword-based search. 
	Design process details are given in \S~\ref{ssec:exp1} and \S~\ref{ssec:exp2} and the final products are in Figures~\ref{fig:product_lara}~and ~\ref{fig:product_butterfly}

	\subsubsection{Desert Monster}
	\label{ssec:exp1}
\begin{figure}[h]
		\centering
		\begin{tabular}[t]{ccc}
			\includegraphics[trim= 0 0 0 0, clip, height=0.171\textwidth]{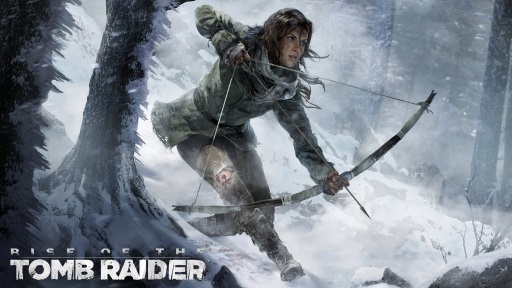} &
			\includegraphics[trim= 0 0 0 0 clip,height=0.171\textwidth]{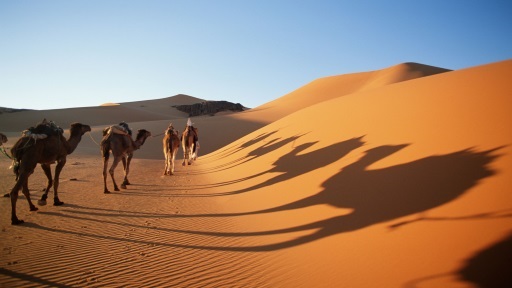} &
			\includegraphics[trim= 0 0 0 0, clip, height=0.171\textwidth]{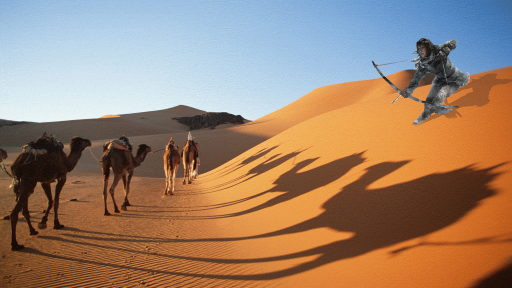} \\
			{\tiny 1} & {\tiny 2} & {\tiny 3}  \\\\
		\end{tabular}
		\begin{tabular}[t]{ccc}
			\includegraphics[trim= 0 0 0 0, clip, height=0.171\textwidth]{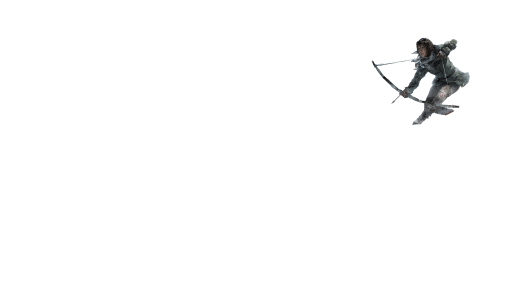} &
			\includegraphics[trim= 0 0 0 0 clip,height=0.171\textwidth]{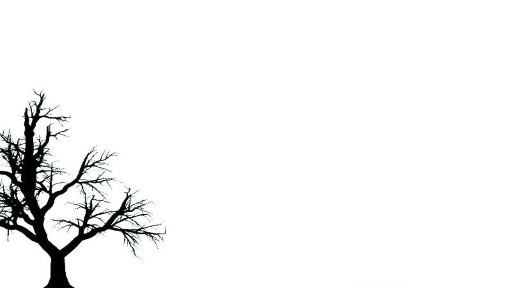} &
			\includegraphics[trim= 0 0 0 0, clip, height=0.171\textwidth]{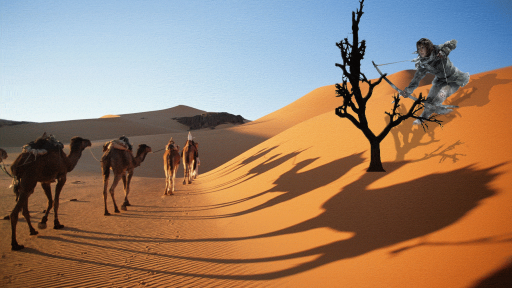} \\
			{\tiny 4} & {\tiny 5} & {\tiny 6}  \\\\
		\end{tabular}
		\begin{tabular}[t]{ccc}
			\includegraphics[trim= 0 0 0 0, clip, height=0.171\textwidth]{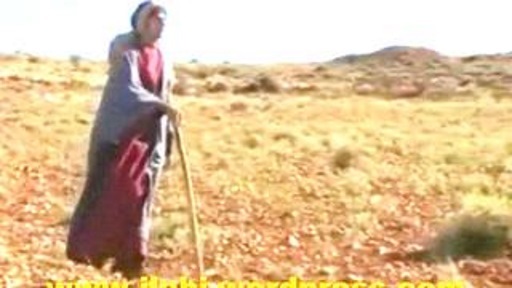} &
			\includegraphics[trim= 0 0 0 0 clip,height=0.171\textwidth]{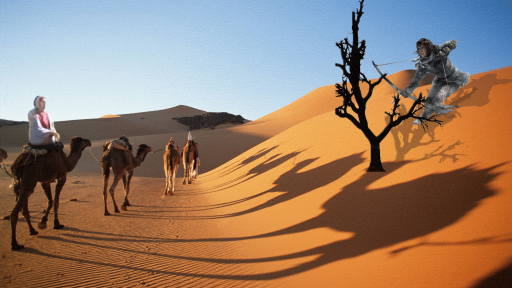} &
			\includegraphics[trim= 0 0 0 0, clip, height=0.171\textwidth]{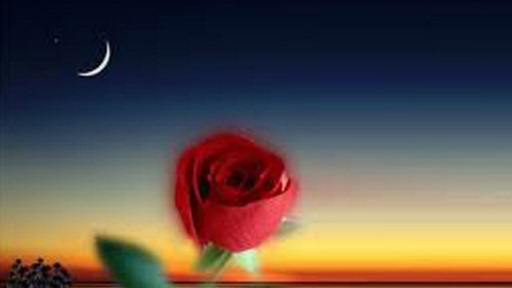} \\
			{\tiny 7} & {\tiny 8} & {\tiny 9} \\\\
		\end{tabular}
		\begin{tabular}[t]{ccc}
			\includegraphics[trim= 0 0 0 0, clip, height=0.171\textwidth]{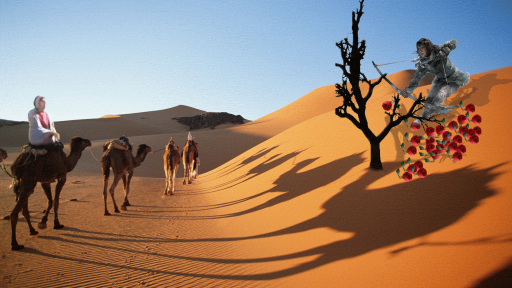} &
			\includegraphics[trim= 0 0 0 0 clip,height=0.171\textwidth]{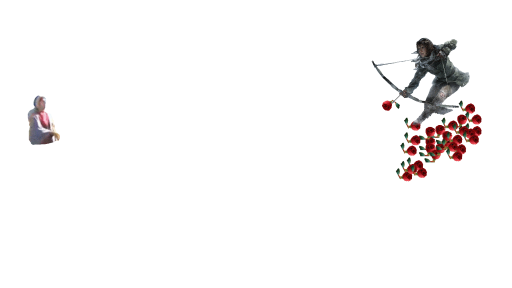} &
			\includegraphics[trim= 0 0 0 0, clip, height=0.171\textwidth]{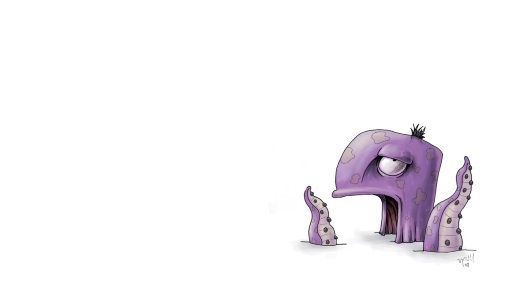} \\
			{\tiny 10} & {\tiny 11} & {\tiny 12}  \\\\
		\end{tabular}	
		\caption{Desert Monster}
		\label{fig:experiment1}
	\end{figure}

	\noindent  
	The evolution of the design is depicted in Figure~\ref{fig:experiment1}.   The user separately queried with the keywords {\sl Lara Croft} and {\sl desert}. Among retrievals to his  keyword-based hunts, he
	picked a Tomb Raider scene  (1) and a desert image (2). After cut-transform-paste operations, he inserted mirrored Lara Croft  on the desert image, yielding (3). In the next step, he used the cut Lara Croft (4)  as a query image for image-based hunt. Among the retrievals came a bush-like black and white tree (5). 
	He pasted it  on the collage (6) and then used the new collage  for the next query that returned a walking nomad woman (7). He cut the woman's upper body  and pasted it in a sitting position on one of the camels of the desert scene (8). He then used the active collage for a new query. Among the retrievals, he picked  a red desert rose (9) and pasted its 
	multiple copies on the collage (10). 
	Then he formed a new query image by removing the background desert image from the collage (11). 
	Quite an unexpected outcome, a purple cartoon creature, squid appeared. Wonderfully surprised, he simply placed this creature on his last collage, naming his design as {\sl Desert Monster}.

\begin{figure}[hbt]
		\centering
		\begin{tabular}[c]{cc}
			\fbox{\includegraphics[height=0.19\linewidth]{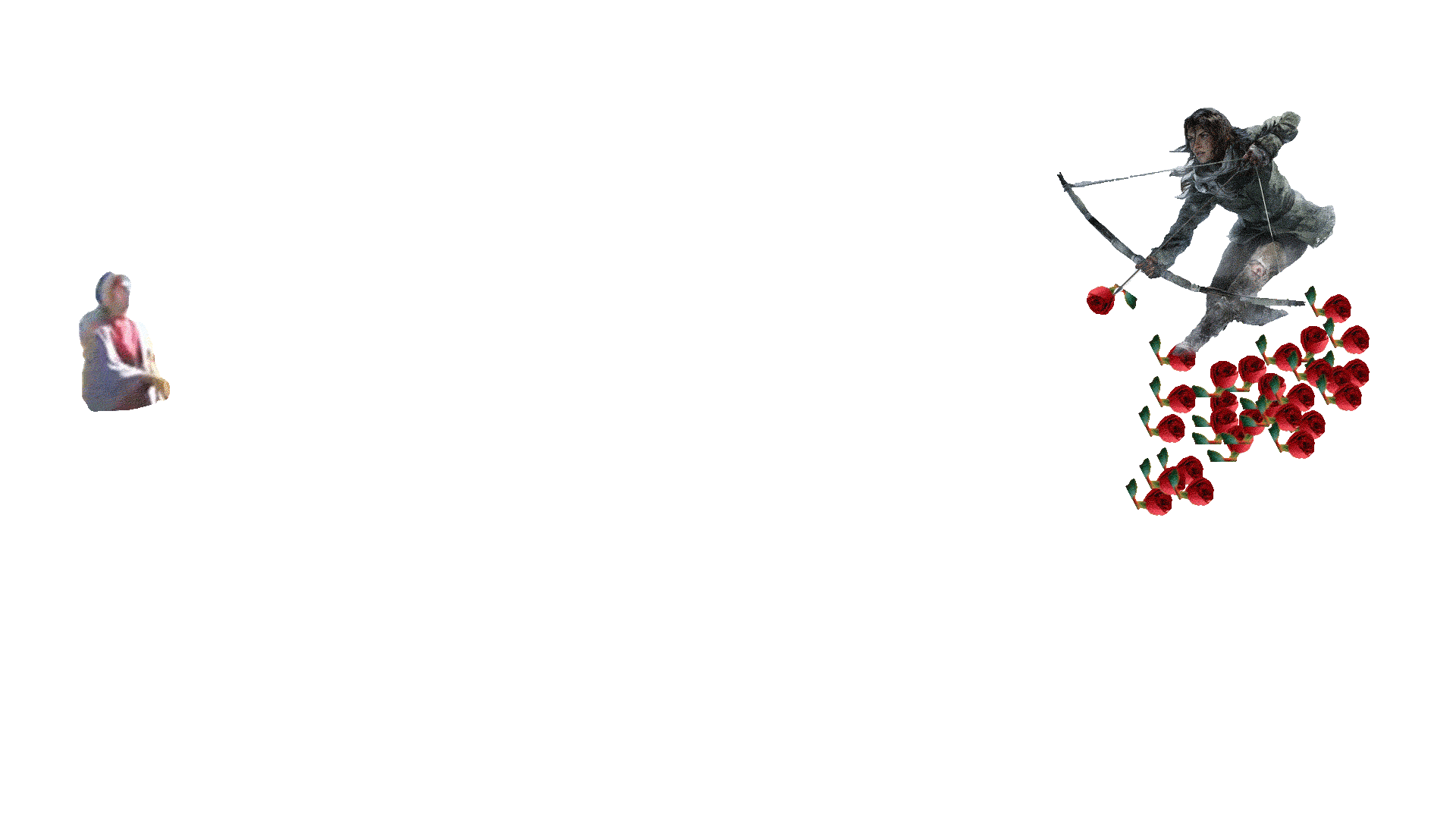}} &
			\includegraphics[height=0.15\linewidth]{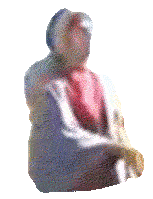}
			\includegraphics[angle=180,origin=c,height=0.15\linewidth]{figures/nomadwomancutted.png}
			\includegraphics[height=0.15\linewidth]{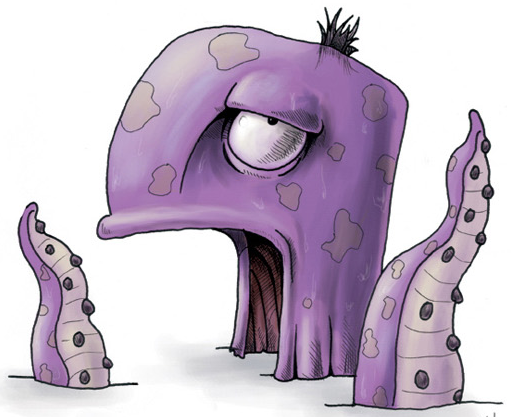}\\  
			(a) & (b)
		\end{tabular}	
		\caption{Retrieval of the purple squid}
		\label{fig:experiment1_part}
	\end{figure}

	The squid popped up in response to an image-based query shown in Figure~\ref{fig:experiment1_part} (a), which is the eleventh step in the process.
	In Figure~\ref{fig:experiment1_part} (b), notice the shawl of the woman: Its color is similar to the color of the squid; its shape, when rotated 180 degrees, is quite similar to the shape of the head of the squid. This might have triggered  the retrieval of the squid. 
	Of course, the user did not pay attention to why this purple cartoon creature popped up among retrievals.  

	\subsection{User Experience: Butterfly}
	\label{ssec:exp2}
\begin{figure}[htb]
		\centering
		\begin{tabular}[t]{cccc}
			\includegraphics[trim= 0 0 0 0, clip, height=0.151\textwidth]{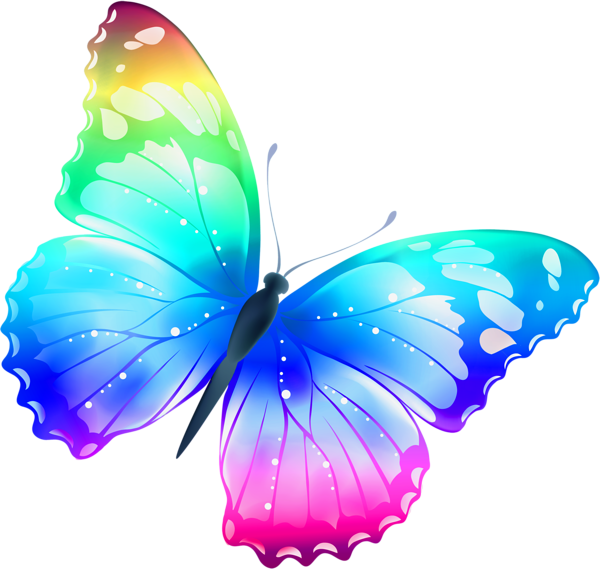} \mbox{  }&
			\includegraphics[trim= 0 0 0 0 clip,height=0.151\textwidth]{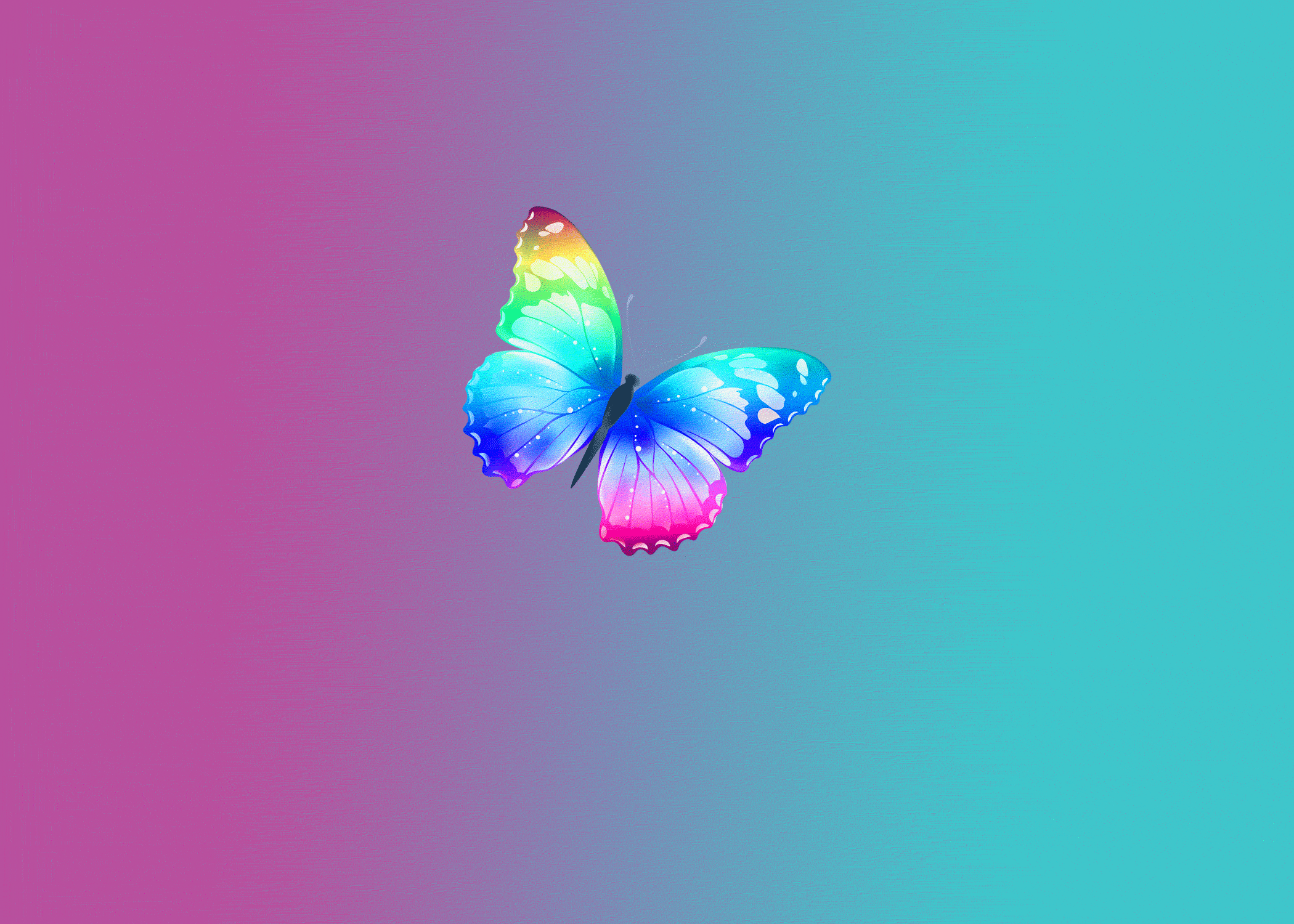} &
			\includegraphics[trim= 0 0 0 0, clip, height=0.151\textwidth]{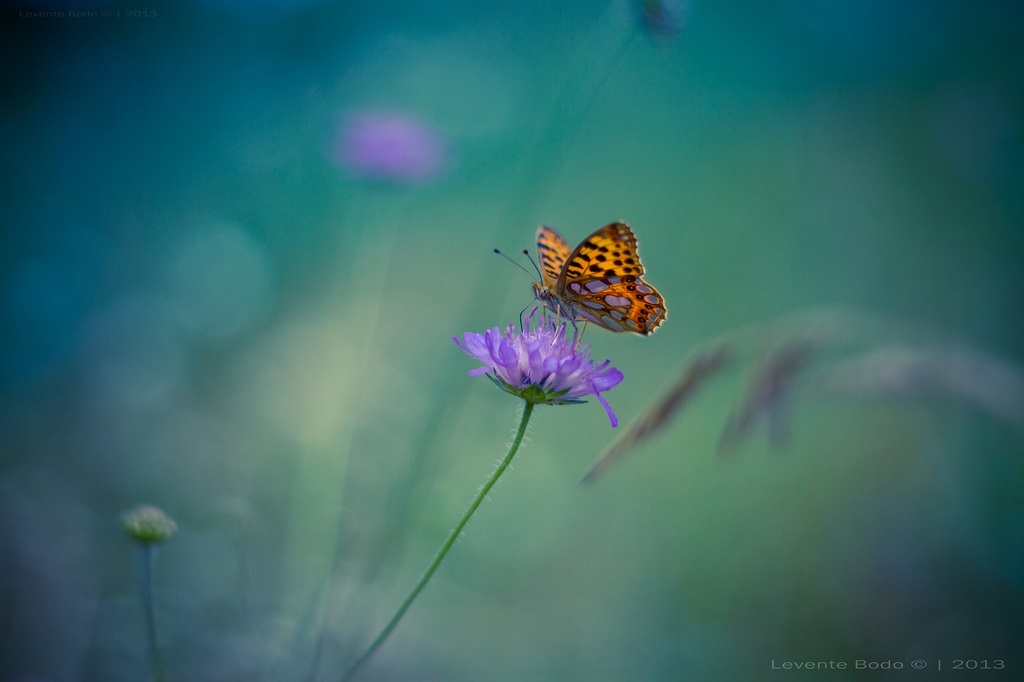} &
			\includegraphics[trim= 0 0 0 0, clip,height=0.151\textwidth]{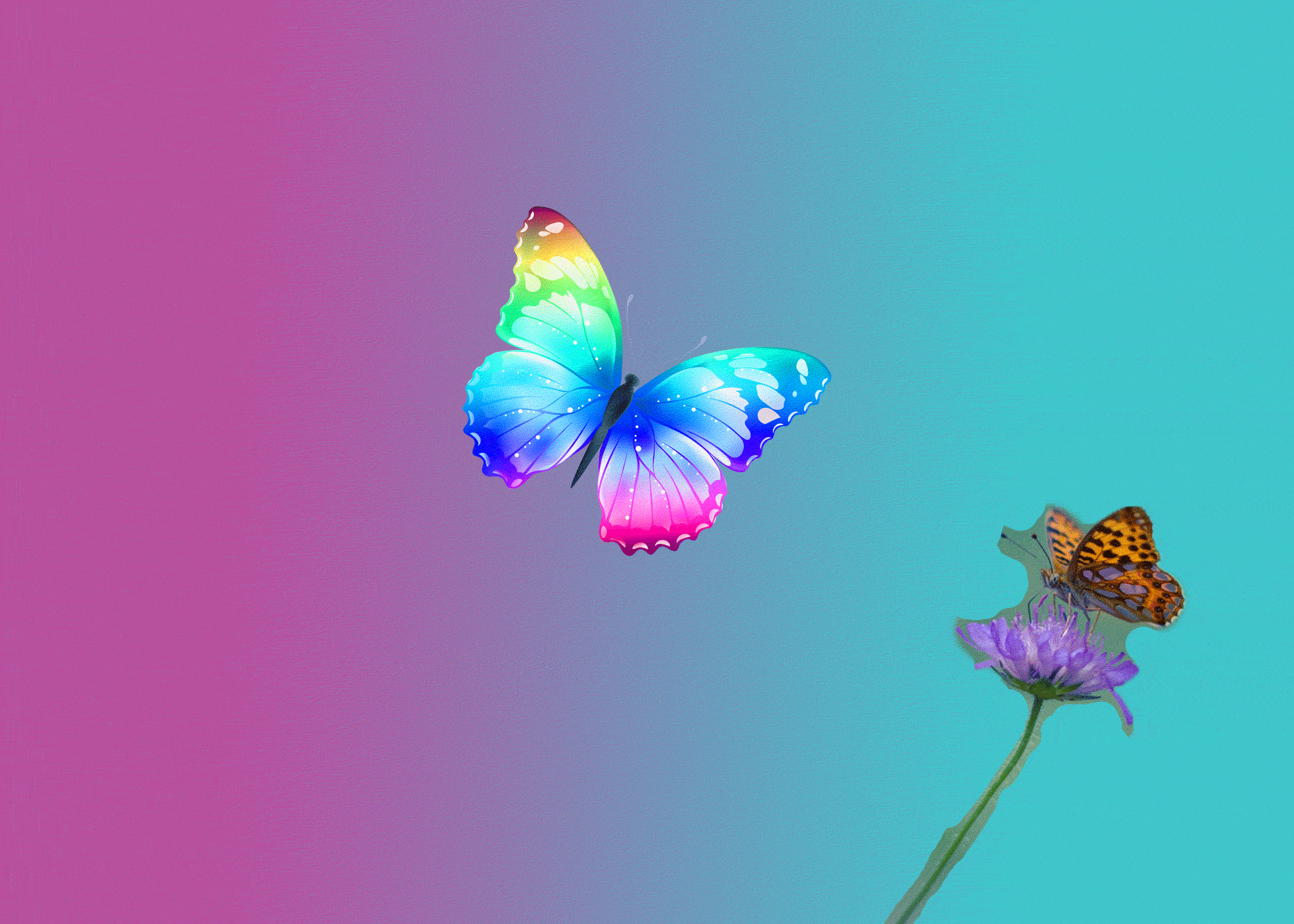}\\
			{\tiny 1} & {\tiny 2} & {\tiny 3} & {\tiny 4} \\\\
		\end{tabular}
		\begin{tabular}[t]{cccc}
			\includegraphics[trim= 0 0 0 0, clip, height=0.153\textwidth]{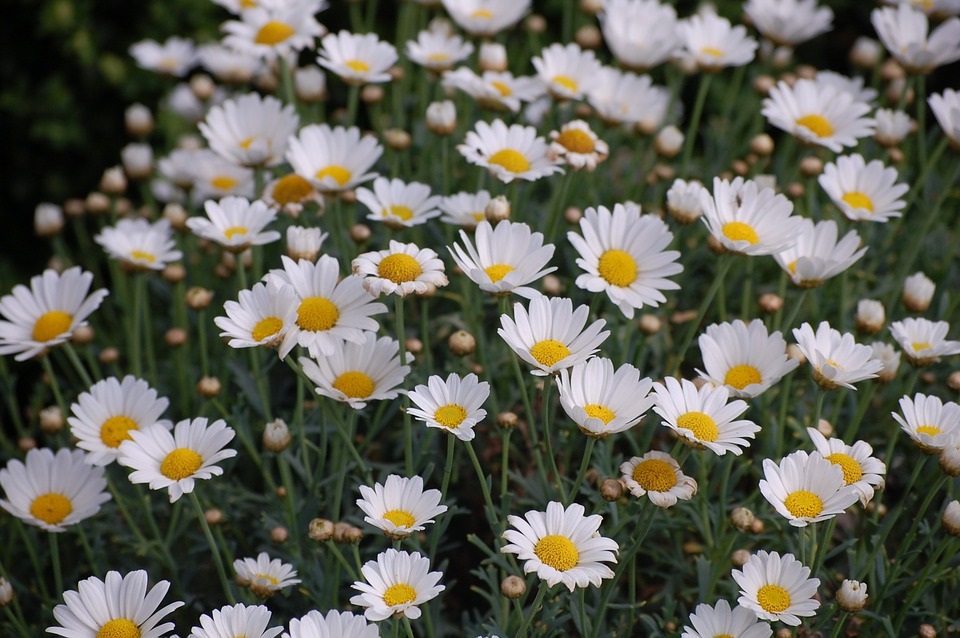} &
			\includegraphics[trim= 0 0 0 0 clip,height=0.153\textwidth]{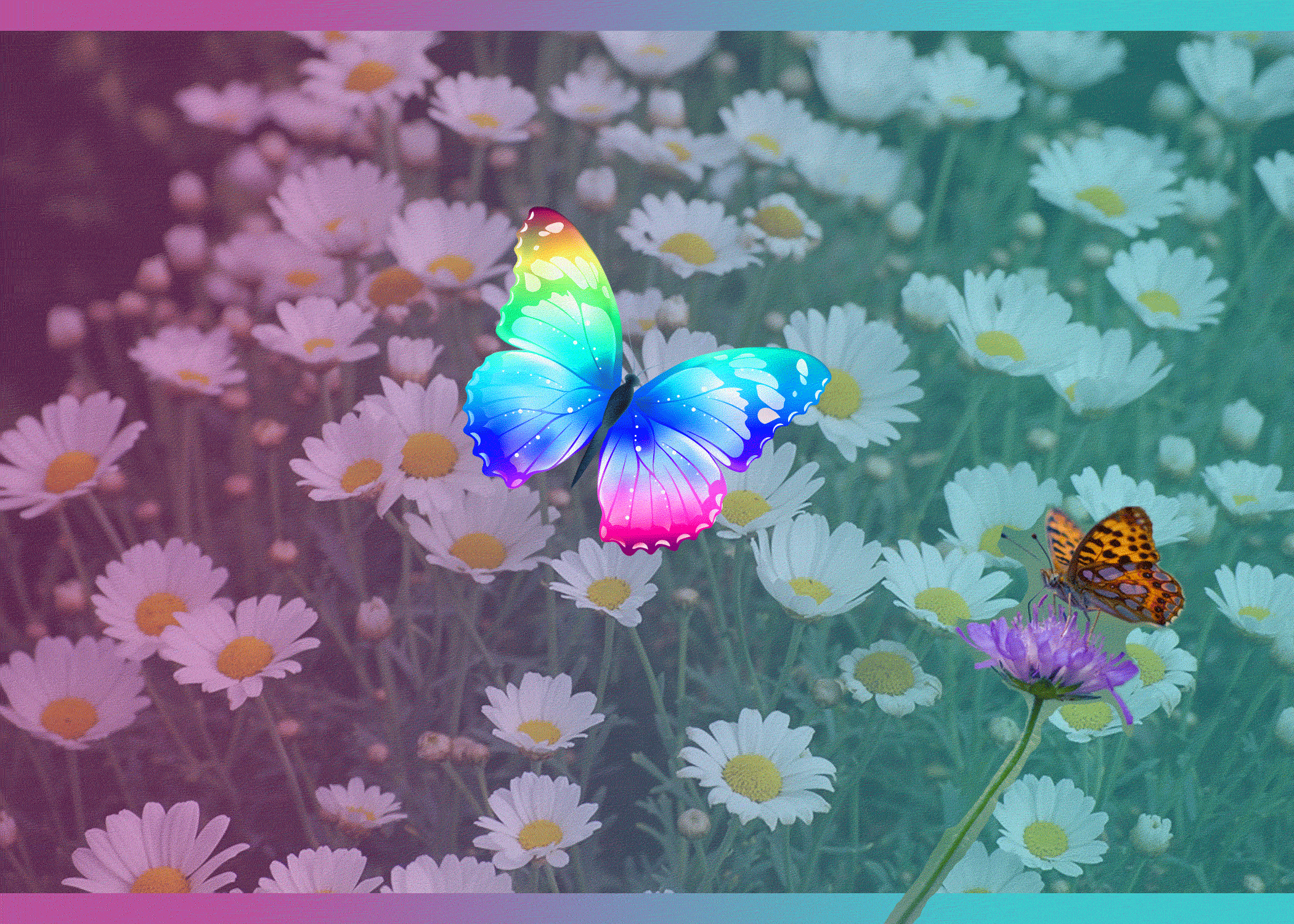} &
			\includegraphics[trim= 0 0 0 0, clip, height=0.153\textwidth]{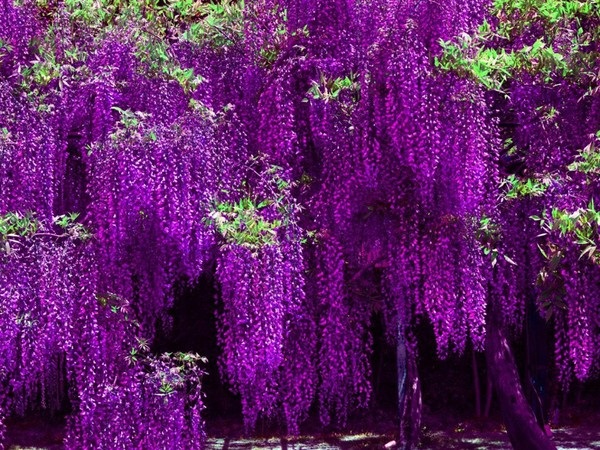} &
			\includegraphics[trim= 0 0 0 0, clip,height=0.153\textwidth]{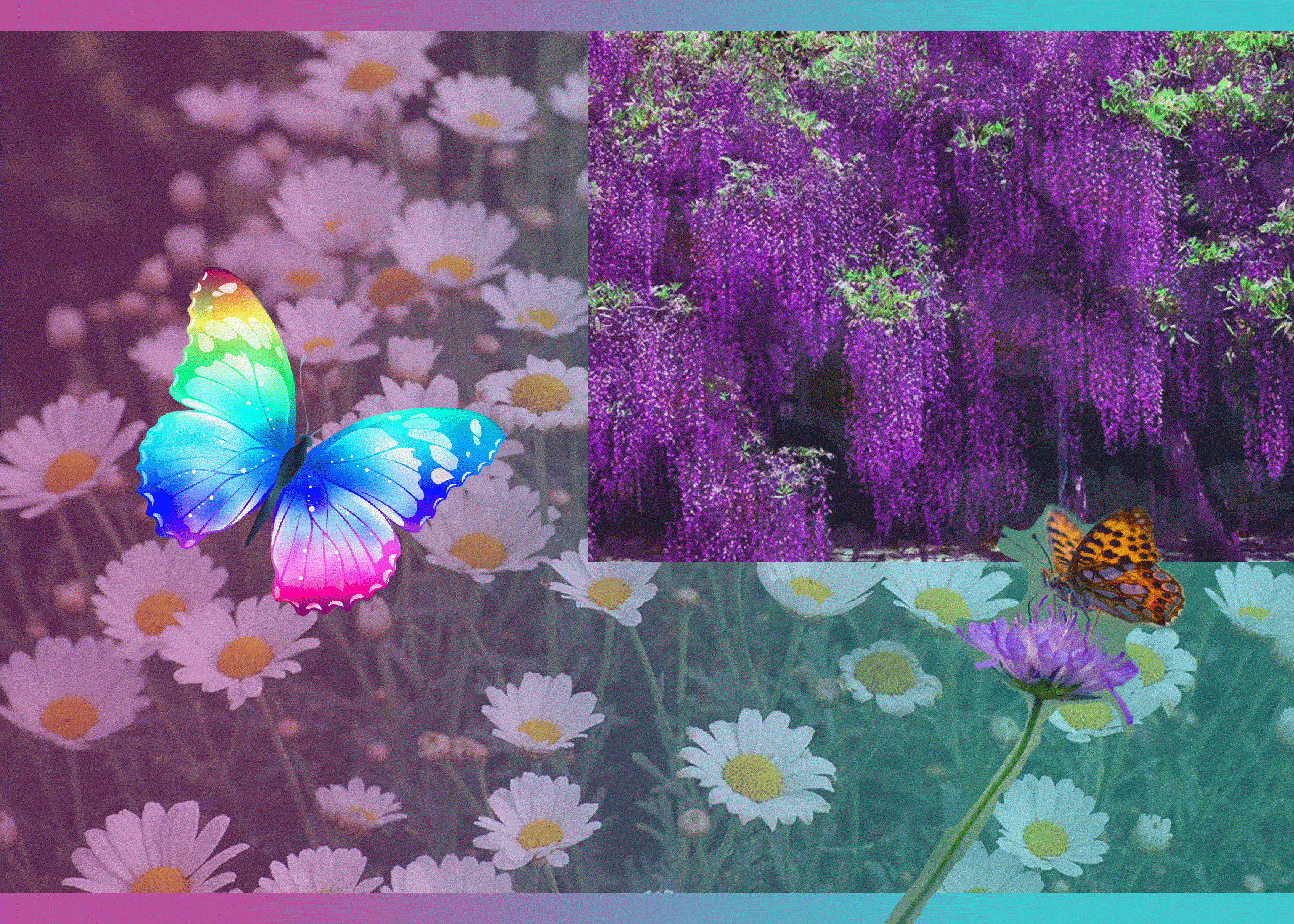}\\
			{\tiny 5} & {\tiny 6} & {\tiny 7} & {\tiny 8} \\\\
		\end{tabular}
		\begin{tabular}[t]{cccc}
			\includegraphics[trim= 0 0 0 0, clip, height=0.155\textwidth]{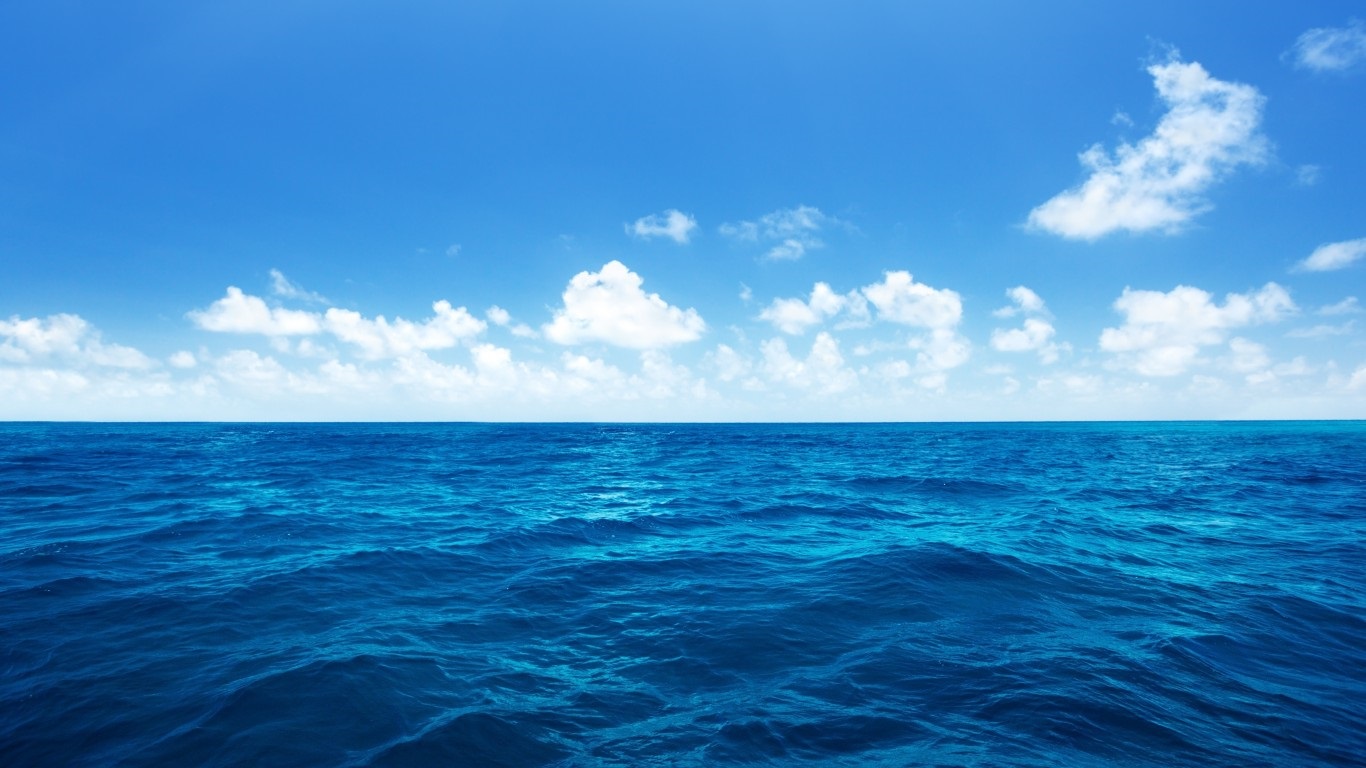} &
			\includegraphics[trim= 0 0 0 0 clip,height=0.155\textwidth]{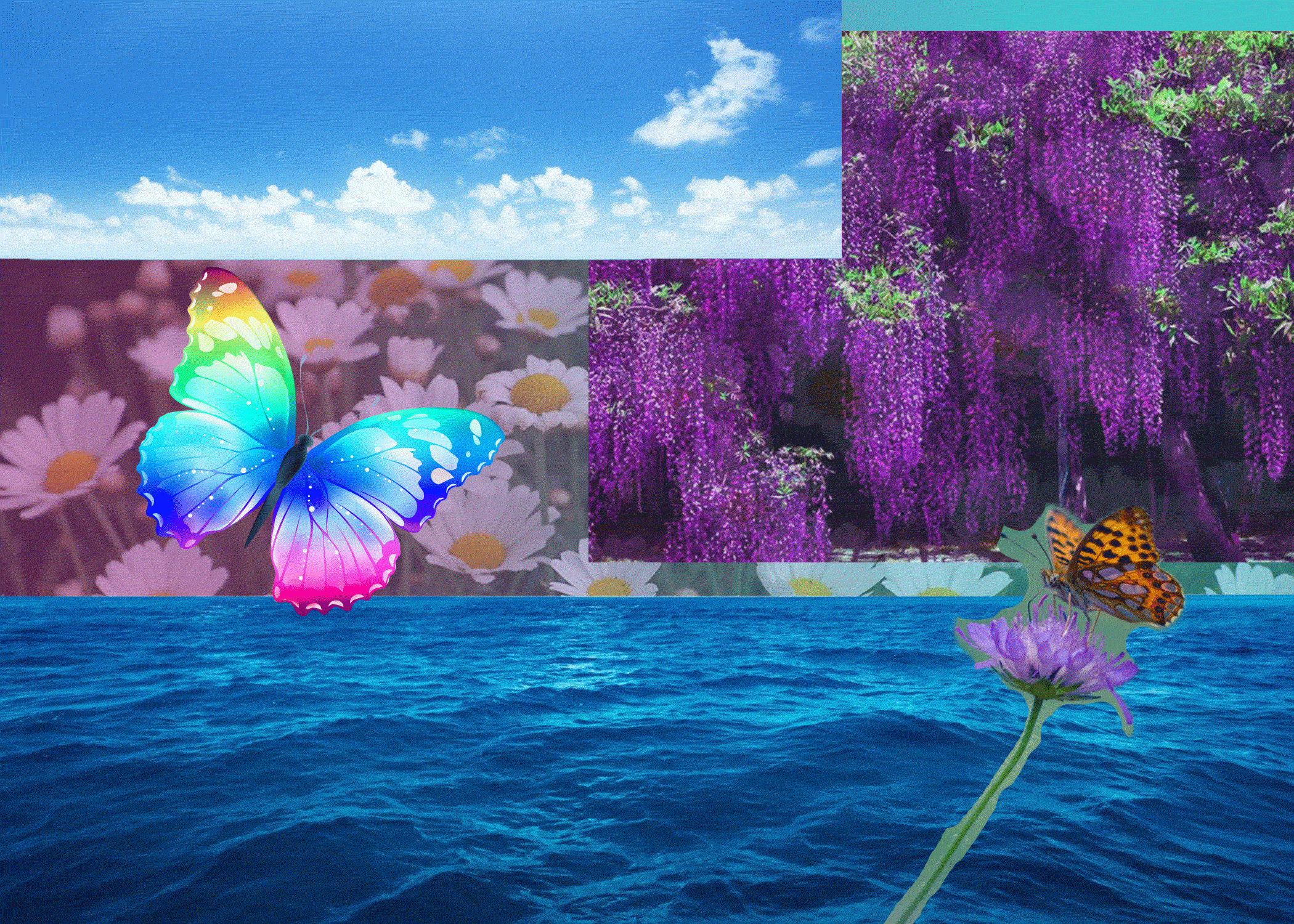} &
			\includegraphics[trim= 0 0 0 0, clip, height=0.155\textwidth]{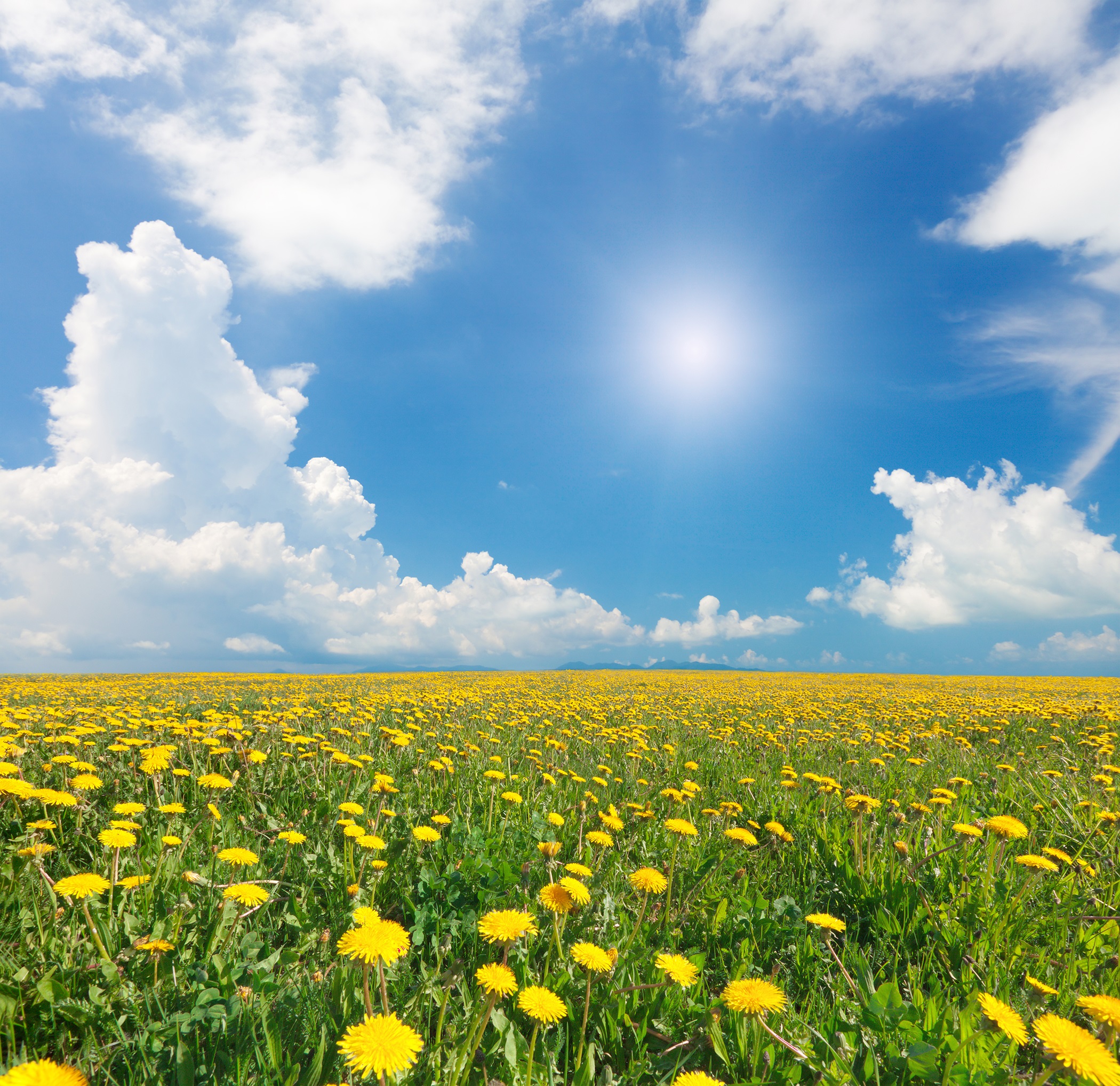} &
			\includegraphics[trim= 0 0 0 0, clip,height=0.155\textwidth]{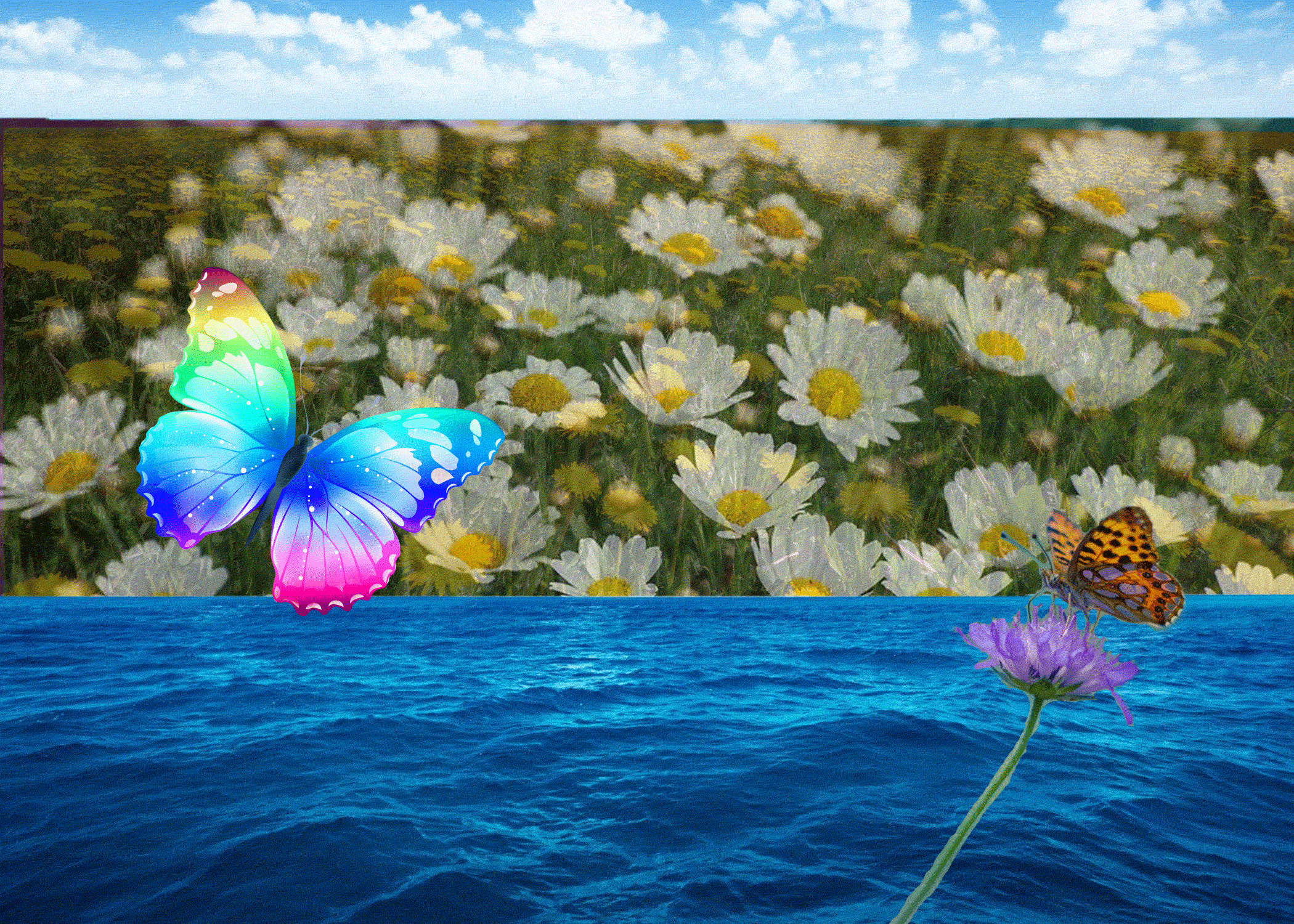}\\
			{\tiny 9} & {\tiny 10} & {\tiny 11} & {\tiny 12} \\\\
		\end{tabular}
		\begin{tabular}[t]{cccc}
			\includegraphics[trim= 0 0 0 0, clip, height=0.158\textwidth]{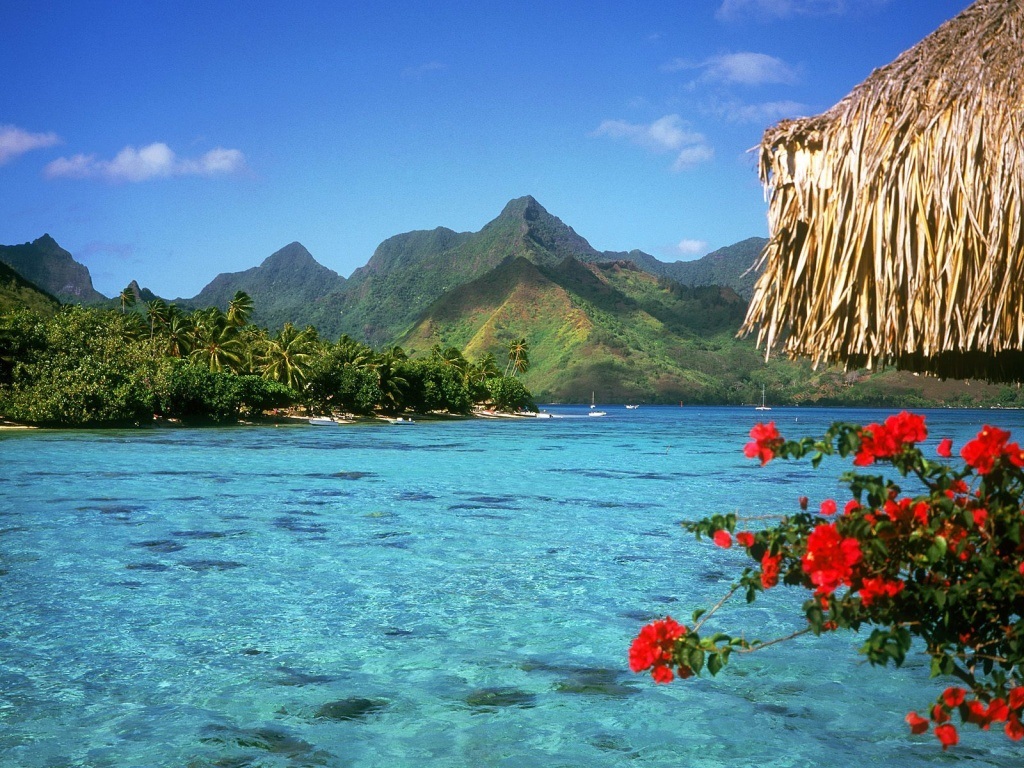} &
			\includegraphics[trim= 0 0 0 0 clip,height=0.158\textwidth]{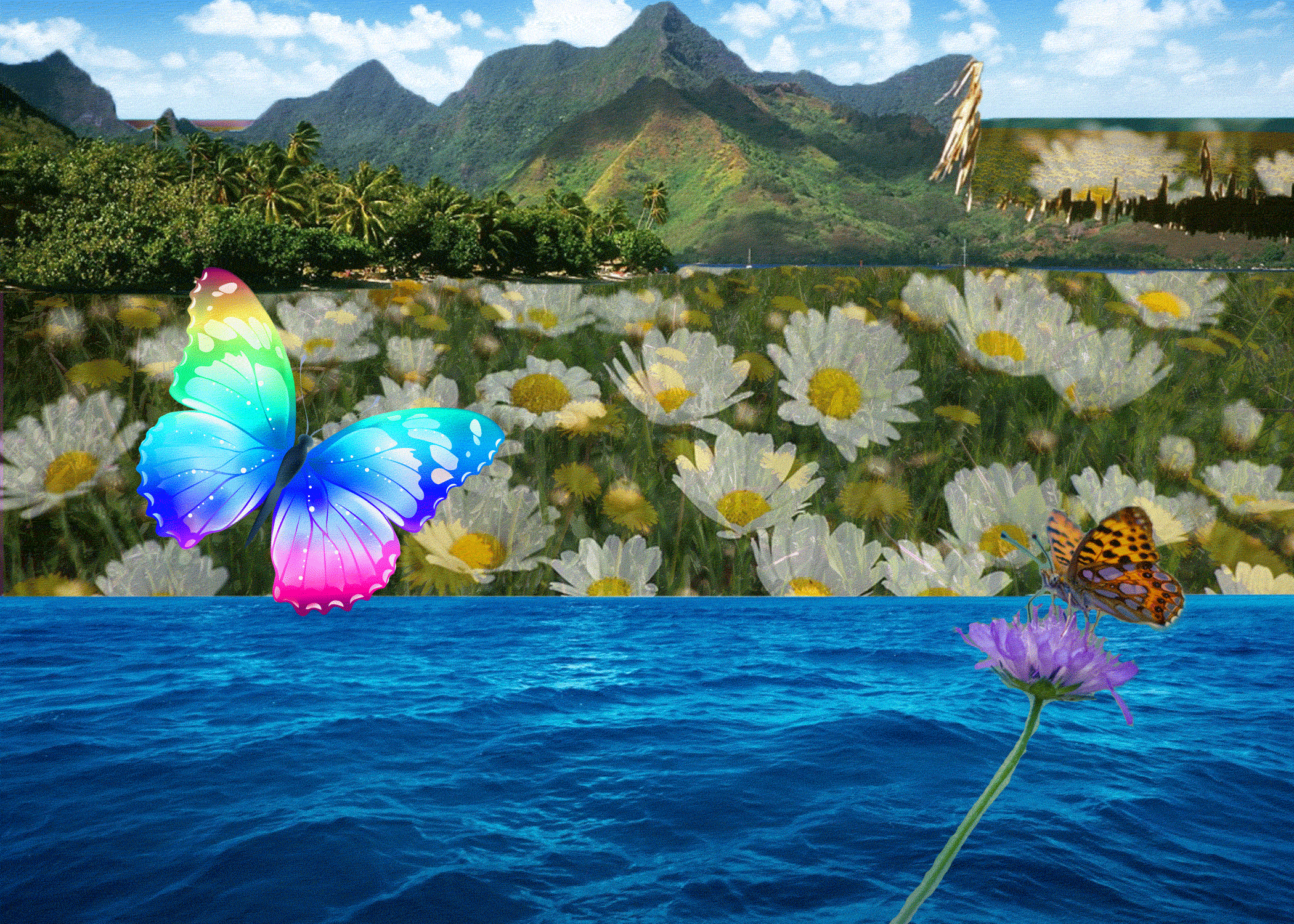} &
			\includegraphics[trim= 0 0 0 0, clip, height=0.158\textwidth]{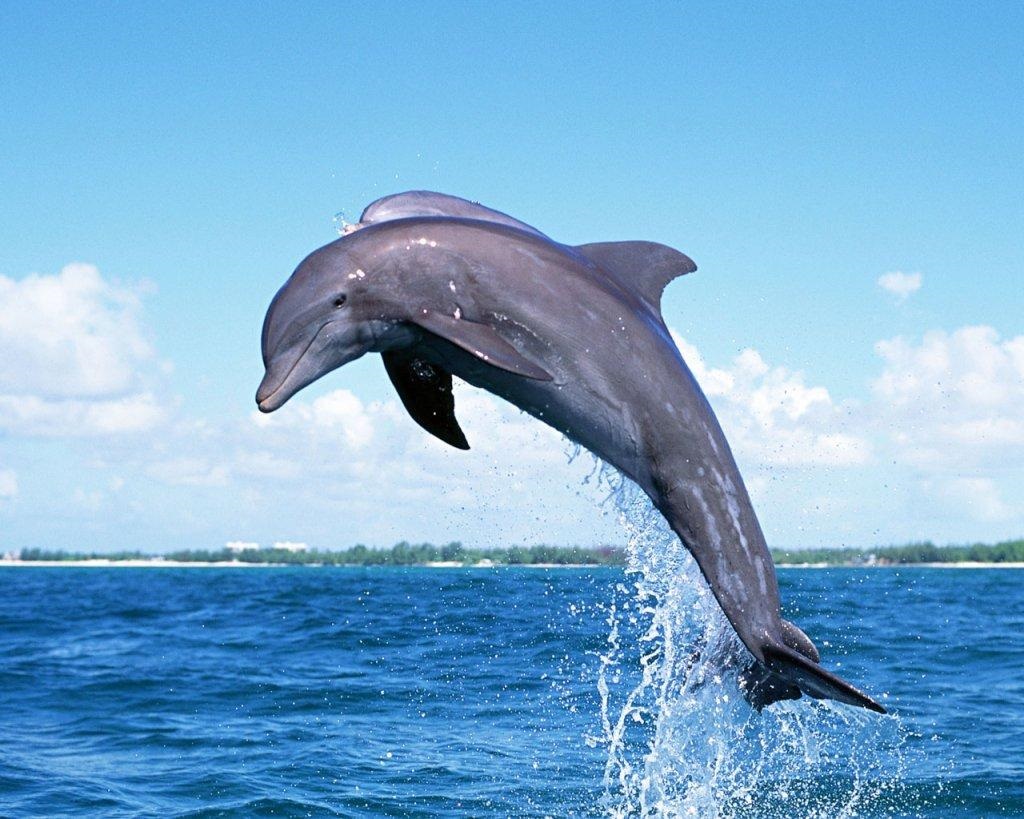} &
			\includegraphics[trim= 0 0 0 0, clip,height=0.158\textwidth]{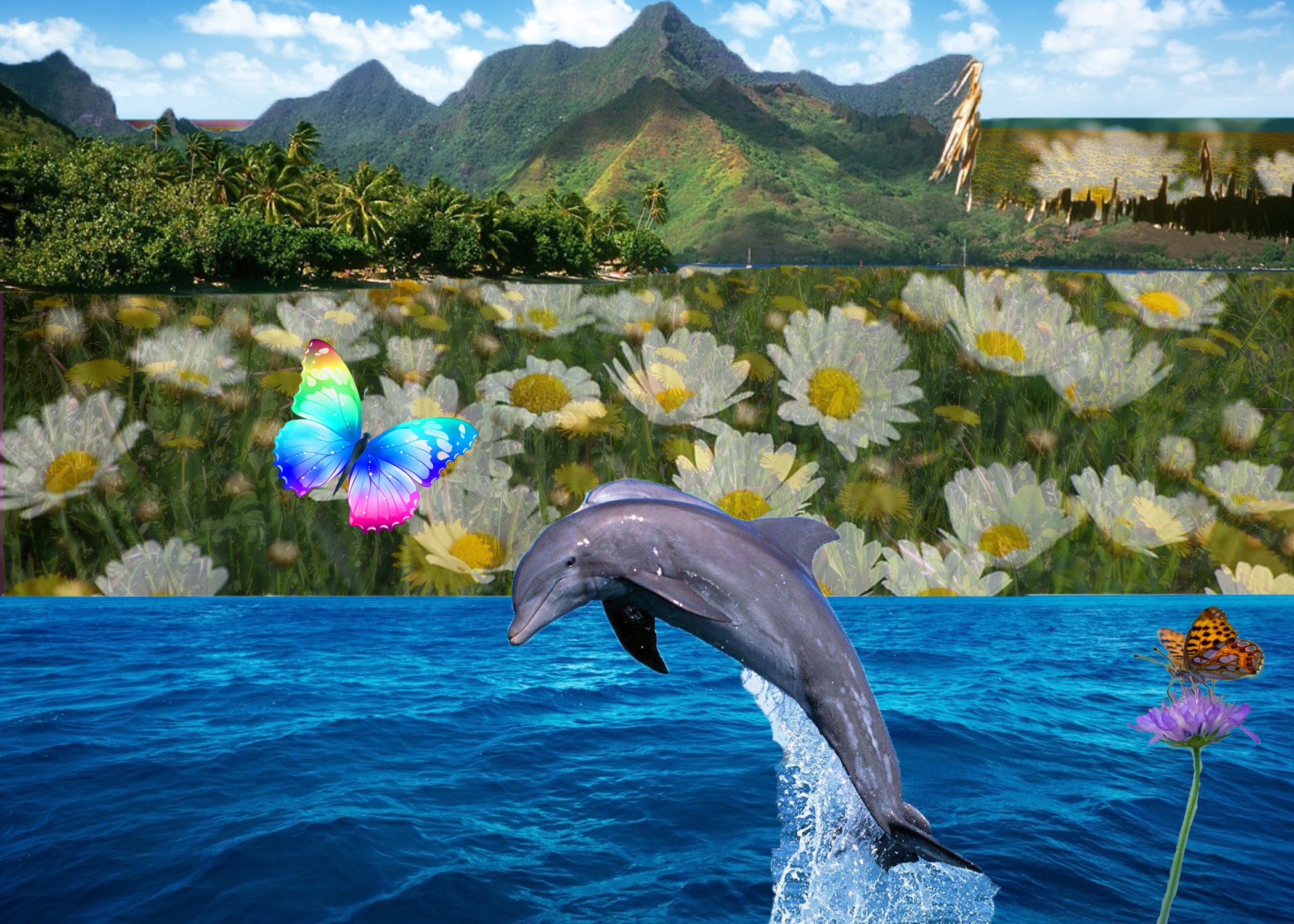}\\
			{\tiny 13} & {\tiny 14} & {\tiny 15} & {\tiny 16} \\\\
		\end{tabular}	
		\caption{Butterfly}
		\label{fig:experiment2}
	\end{figure}

	\noindent 
	The evolution of the design is depicted in Figure~\ref{fig:experiment2}. The user started with a keyword-based search using {\sl butterfly} as the keyword. 
	Among retrievals, he picked a technicolor butterfly (1). Using Photoshop tools, he constructed a background image mimicking the colors of the butterfly on which the butterfly is pasted (2). He then queried using the active collage in (2) and picked  the image in (3): a monarch on a purple flower. After cut-transform-paste, he used the modified collage in (4) for another image-based query. He integrated a query result, daisy field in (5),  as a background and used the new collage  for the next image-based query which retrieved a purple flower tree forest image (7). He pasted (7)   on the current collage and used the resulting image (8) for the next query. Unsatisfied with the retrievals, he  switched to a keyword-based search to introduce some blue tone. He used {\sl sea} as the keyword and picked the sea image shown in (9). He made a new keyword-based search using {\sl air} as the keyword. Nonetheless, he decided to use the previous result of the {\sl sea}keyword query shown in (9) query to cut and paste both the sea and sky (10). 
	He continued with an image-based search using the active collage (10) and retrieved a field of yellow dandelion flowers (11).

	He formed a new background by removing the upper transparent layer, making the daisy field layer transparent and adding the dandelion field as a bottom layer as shown in (12). He also removed the purple tree. The resulting collage (12) is used for another image-based search which returned the landscape in (13).  
	From (13), he  cut  the mountain region and pasted it on the collage forming (14). 
	Image-based search with the active collage (14) retrieved a dolphin (15). He placed the dolphin as the last element to his collage (16) and finalizing the content after removing the initial technicolor butterfly.  
	Notice how some removed elements (initial butterfly, the purple tree) shaped the design and contributed to the color harmony even though they are not included in the final product.

	\section{Conclusion}
	\label{sec:conc}
	\noindent 
	Reported user experiences  showed that Inspiration Hunter  supports reuse of existing material (crowd-generated big data on the web) in new and unexpected ways. Thus, it is  a step towards data-enabled, iterative, imitative, open and collaborative  creativity in picture design. 

	A future improvement is to place server side operations on a cloud to boost reusability.  This is especially useful for technically-apt platform users who wish to add their custom features, without needing to re-develop existing services. 
	Server side maintenance and management should be assigned to cloud to reduce the effort to maintain the system. 

	\FloatBarrier
	\clearpage
	\bibliographystyle{apacite}
	%\bibliography{journal_ST_final_July152020}

\begin{thebibliography}{}
		
		\bibitem [\protect \citeauthoryear {%
			El-Zanfaly%
		}{%
			El-Zanfaly%
		}{%
			{\protect \APACyear {2015}}%
		}]{%
			ElZanfaly}
		\APACinsertmetastar {%
			ElZanfaly}%
		\begin{APACrefauthors}%
			El-Zanfaly, D.%
		\end{APACrefauthors}%
		\unskip\
		\newblock
		\APACrefYearMonthDay{2015}{}{}.
		\newblock
		{\BBOQ}\APACrefatitle {{[I3] Imitation, Iteration and Improvisation: Embodied
				interaction in making and learning}} {{[I3] Imitation, Iteration and
				Improvisation: Embodied interaction in making and learning}}.{\BBCQ}
		\newblock
		\APACjournalVolNumPages{Design Studies}{41}{}{79--109}.
		\newblock
		\begin{APACrefDOI} \doi{10.1016/j.destud.2015.09.002} \end{APACrefDOI}
		\PrintBackRefs{\CurrentBib}
		
		\bibitem [\protect \citeauthoryear {%
			Englard%
		}{%
			Englard%
		}{%
			{\protect \APACyear {2016}}%
		}]{%
			Waffle}
		\APACinsertmetastar {%
			Waffle}%
		\begin{APACrefauthors}%
			Englard, B.%
		\end{APACrefauthors}%
		\unskip\
		\newblock
		\APACrefYearMonthDay{2016}{}{}.
		\newblock
		\APACrefbtitle {{Waffle}.} {{Waffle}.}
		\newblock
		\begin{APACrefURL} \url{https://github.com/benglard/waffle} \end{APACrefURL}
		\PrintBackRefs{\CurrentBib}
		
		\bibitem [\protect \citeauthoryear {%
			Ensari%
			\ \BBA {} {\"{O}}zkar%
		}{%
			Ensari%
			\ \BBA {} {\"{O}}zkar%
		}{%
			{\protect \APACyear {2018}}%
		}]{%
			ensari_ozkar_2018}
		\APACinsertmetastar {%
			ensari_ozkar_2018}%
		\begin{APACrefauthors}%
			Ensari, E.%
			\BCBT {}\ \BBA {} {\"{O}}zkar, M.%
		\end{APACrefauthors}%
		\unskip\
		\newblock
		\APACrefYearMonthDay{2018}{}{}.
		\newblock
		{\BBOQ}\APACrefatitle {Shape computations with NURB curves} {Shape computations
			with nurb curves}.{\BBCQ}
		\newblock
		\APACjournalVolNumPages{Artificial Intelligence for Engineering Design,
			Analysis and Manufacturing}{32}{3}{282–294}.
		\newblock
		\begin{APACrefDOI} \doi{10.1017/S0890060417000592} \end{APACrefDOI}
		\PrintBackRefs{\CurrentBib}
		
		\bibitem [\protect \citeauthoryear {%
			Goucher-Lambert%
			\ \BBA {} Cagan%
		}{%
			Goucher-Lambert%
			\ \BBA {} Cagan%
		}{%
			{\protect \APACyear {2019}}%
		}]{%
			Cagan2019}
		\APACinsertmetastar {%
			Cagan2019}%
		\begin{APACrefauthors}%
			Goucher-Lambert, K.%
			\BCBT {}\ \BBA {} Cagan, J.%
		\end{APACrefauthors}%
		\unskip\
		\newblock
		\APACrefYearMonthDay{2019}{}{}.
		\newblock
		{\BBOQ}\APACrefatitle {{Crowdsourcing inspiration: Using crowd generated
				inspirational stimuli to support designer ideation}} {{Crowdsourcing
				inspiration: Using crowd generated inspirational stimuli to support designer
				ideation}}.{\BBCQ}
		\newblock
		\APACjournalVolNumPages{Design Studies}{61}{}{1--29}.
		\newblock
		\begin{APACrefURL}
			\url{https://linkinghub.elsevier.com/retrieve/pii/S0142694X19300018}
		\end{APACrefURL}
		\newblock
		\begin{APACrefDOI} \doi{10.1016/j.destud.2019.01.001} \end{APACrefDOI}
		\PrintBackRefs{\CurrentBib}
		
		\bibitem [\protect \citeauthoryear {%
			Grasl%
			\ \BBA {} Economou%
		}{%
			Grasl%
			\ \BBA {} Economou%
		}{%
			{\protect \APACyear {2013}}%
		}]{%
			Grasl2013}
		\APACinsertmetastar {%
			Grasl2013}%
		\begin{APACrefauthors}%
			Grasl, T.%
			\BCBT {}\ \BBA {} Economou, A.%
		\end{APACrefauthors}%
		\unskip\
		\newblock
		\APACrefYearMonthDay{2013}{}{}.
		\newblock
		{\BBOQ}\APACrefatitle {From Topologies to Shapes: Parametric Shape Grammars
			Implemented by Graphs} {From topologies to shapes: Parametric shape grammars
			implemented by graphs}.{\BBCQ}
		\newblock
		\APACjournalVolNumPages{Environment and Planning B: Planning and
			Design}{}{5}{905–922}.
		\PrintBackRefs{\CurrentBib}
		
		\bibitem [\protect \citeauthoryear {%
			G{\"{u}}rsoy%
			\ \BBA {} {\"{O}}zkar%
		}{%
			G{\"{u}}rsoy%
			\ \BBA {} {\"{O}}zkar%
		}{%
			{\protect \APACyear {2015}}%
		}]{%
			GursoyOzkar}
		\APACinsertmetastar {%
			GursoyOzkar}%
		\begin{APACrefauthors}%
			G{\"{u}}rsoy, B.%
			\BCBT {}\ \BBA {} {\"{O}}zkar, M.%
		\end{APACrefauthors}%
		\unskip\
		\newblock
		\APACrefYearMonthDay{2015}{}{}.
		\newblock
		{\BBOQ}\APACrefatitle {{Visualizing making: Shapes, materials, and actions}}
		{{Visualizing making: Shapes, materials, and actions}}.{\BBCQ}
		\newblock
		\APACjournalVolNumPages{Design Studies}{41}{}{29--50}.
		\newblock
		\begin{APACrefDOI} \doi{10.1016/j.destud.2015.08.007} \end{APACrefDOI}
		\PrintBackRefs{\CurrentBib}
		
		\bibitem [\protect \citeauthoryear {%
			Harrison%
			, Earl%
			\BCBL {}\ \BBA {} Eckert%
		}{%
			Harrison%
			\ \protect \BOthers {.}}{%
			{\protect \APACyear {2015}}%
		}]{%
			Harrison}
		\APACinsertmetastar {%
			Harrison}%
		\begin{APACrefauthors}%
			Harrison, L.%
			, Earl, C.%
			\BCBL {}\ \BBA {} Eckert, C.%
		\end{APACrefauthors}%
		\unskip\
		\newblock
		\APACrefYearMonthDay{2015}{}{}.
		\newblock
		{\BBOQ}\APACrefatitle {{Exploratory making: Shape, structure and motion}}
		{{Exploratory making: Shape, structure and motion}}.{\BBCQ}
		\newblock
		\APACjournalVolNumPages{Design Studies}{41}{A}{51--78}.
		\newblock
		\begin{APACrefDOI} \doi{10.1016/j.destud.2015.08.003} \end{APACrefDOI}
		\PrintBackRefs{\CurrentBib}
		
		\bibitem [\protect \citeauthoryear {%
			Johnson%
		}{%
			Johnson%
		}{%
			{\protect \APACyear {2015}}%
		}]{%
			Johnson2015}
		\APACinsertmetastar {%
			Johnson2015}%
		\begin{APACrefauthors}%
			Johnson, J.%
		\end{APACrefauthors}%
		\unskip\
		\newblock
		\APACrefYearMonthDay{2015}{}{}.
		\newblock
		\APACrefbtitle {neural-style.} {neural-style.}
		\newblock
		\begin{APACrefURL} \url{https://github.com/jcjohnson/neural-style}
		\end{APACrefURL}
		\PrintBackRefs{\CurrentBib}
		
		\bibitem [\protect \citeauthoryear {%
			Keles%
			, {\"{O}}zkar%
			\BCBL {}\ \BBA {} Tari%
		}{%
			Keles%
			\ \protect \BOthers {.}}{%
			{\protect \APACyear {2010}}%
		}]{%
			Keles2010}
		\APACinsertmetastar {%
			Keles2010}%
		\begin{APACrefauthors}%
			Keles, H\BPBI Y.%
			, {\"{O}}zkar, M.%
			\BCBL {}\ \BBA {} Tari, S.%
		\end{APACrefauthors}%
		\unskip\
		\newblock
		\APACrefYearMonthDay{2010}{}{}.
		\newblock
		{\BBOQ}\APACrefatitle {Embedding shapes without predefined parts} {Embedding
			shapes without predefined parts}.{\BBCQ}
		\newblock
		\APACjournalVolNumPages{Environment and Planning B: Planning and
			Design}{37}{}{664--681}.
		\PrintBackRefs{\CurrentBib}
		
		\bibitem [\protect \citeauthoryear {%
			Keles%
			, {\"{O}}zkar%
			\BCBL {}\ \BBA {} Tari%
		}{%
			Keles%
			\ \protect \BOthers {.}}{%
			{\protect \APACyear {2012}}%
		}]{%
			Keles2012}
		\APACinsertmetastar {%
			Keles2012}%
		\begin{APACrefauthors}%
			Keles, H\BPBI Y.%
			, {\"{O}}zkar, M.%
			\BCBL {}\ \BBA {} Tari, S.%
		\end{APACrefauthors}%
		\unskip\
		\newblock
		\APACrefYearMonthDay{2012}{}{}.
		\newblock
		{\BBOQ}\APACrefatitle {Weighted shapes for embedding perceived wholes}
		{Weighted shapes for embedding perceived wholes}.{\BBCQ}
		\newblock
		\APACjournalVolNumPages{Environment and Planning B: Planning and
			Design}{39}{}{360--375}.
		\PrintBackRefs{\CurrentBib}
		
		\bibitem [\protect \citeauthoryear {%
			Knight%
			\ \BBA {} Vardouli%
		}{%
			Knight%
			\ \BBA {} Vardouli%
		}{%
			{\protect \APACyear {2015}}%
		}]{%
			KnightVardouli}
		\APACinsertmetastar {%
			KnightVardouli}%
		\begin{APACrefauthors}%
			Knight, T.%
			\BCBT {}\ \BBA {} Vardouli, T.%
		\end{APACrefauthors}%
		\unskip\
		\newblock
		\APACrefYearMonthDay{2015}{}{}.
		\newblock
		{\BBOQ}\APACrefatitle {Making grammars: From computing with shapes to computing
			with things} {Making grammars: From computing with shapes to computing with
			things}.{\BBCQ}
		\newblock
		\APACjournalVolNumPages{Design Studies}{41}{A}{1--7}.
		\PrintBackRefs{\CurrentBib}
		
		\bibitem [\protect \citeauthoryear {%
			Prats%
			, Earl%
			, Garner%
			\BCBL {}\ \BBA {} Jowers%
		}{%
			Prats%
			\ \protect \BOthers {.}}{%
			{\protect \APACyear {2006}}%
		}]{%
			prats_earl_garner_jowers_2006}
		\APACinsertmetastar {%
			prats_earl_garner_jowers_2006}%
		\begin{APACrefauthors}%
			Prats, M.%
			, Earl, C.%
			, Garner, S.%
			\BCBL {}\ \BBA {} Jowers, I.%
		\end{APACrefauthors}%
		\unskip\
		\newblock
		\APACrefYearMonthDay{2006}{}{}.
		\newblock
		{\BBOQ}\APACrefatitle {Shape exploration of designs in a style: Toward
			generation of product designs} {Shape exploration of designs in a style:
			Toward generation of product designs}.{\BBCQ}
		\newblock
		\APACjournalVolNumPages{Artificial Intelligence for Engineering Design,
			Analysis and Manufacturing}{20}{3}{201–215}.
		\newblock
		\begin{APACrefDOI} \doi{10.1017/S0890060406060173} \end{APACrefDOI}
		\PrintBackRefs{\CurrentBib}
		
		\bibitem [\protect \citeauthoryear {%
			Sarkar%
			\ \BBA {} Chakrabarti%
		}{%
			Sarkar%
			\ \BBA {} Chakrabarti%
		}{%
			{\protect \APACyear {2008}}%
		}]{%
			sarkar_chakrabarti_AIEDAM}
		\APACinsertmetastar {%
			sarkar_chakrabarti_AIEDAM}%
		\begin{APACrefauthors}%
			Sarkar, P.%
			\BCBT {}\ \BBA {} Chakrabarti, A.%
		\end{APACrefauthors}%
		\unskip\
		\newblock
		\APACrefYearMonthDay{2008}{}{}.
		\newblock
		{\BBOQ}\APACrefatitle {The effect of representation of triggers on design
			outcomes} {The effect of representation of triggers on design
			outcomes}.{\BBCQ}
		\newblock
		\APACjournalVolNumPages{Artificial Intelligence for Engineering Design,
			Analysis and Manufacturing}{22}{2}{101–116}.
		\newblock
		\begin{APACrefDOI} \doi{10.1017/S0890060408000073} \end{APACrefDOI}
		\PrintBackRefs{\CurrentBib}
		
		\bibitem [\protect \citeauthoryear {%
			Sawyer%
		}{%
			Sawyer%
		}{%
			{\protect \APACyear {2018}}%
		}]{%
			Sawyer}
		\APACinsertmetastar {%
			Sawyer}%
		\begin{APACrefauthors}%
			Sawyer, R\BPBI K.%
		\end{APACrefauthors}%
		\unskip\
		\newblock
		\APACrefYearMonthDay{2018}{}{}.
		\newblock
		{\BBOQ}\APACrefatitle {{How Artists Create: An Empirical Study of MFA Painting
				Students}} {{How Artists Create: An Empirical Study of MFA Painting
				Students}}.{\BBCQ}
		\newblock
		\APACjournalVolNumPages{Journal of Creative Behavior}{52}{2}{127--141}.
		\newblock
		\begin{APACrefDOI} \doi{10.1002/jocb.136} \end{APACrefDOI}
		\PrintBackRefs{\CurrentBib}
		
		\bibitem [\protect \citeauthoryear {%
			Smith%
		}{%
			Smith%
		}{%
			{\protect \APACyear {2010}}%
		}]{%
			GoogleImages}
		\APACinsertmetastar {%
			GoogleImages}%
		\begin{APACrefauthors}%
			Smith, N.%
		\end{APACrefauthors}%
		\unskip\
		\newblock
		\APACrefYearMonthDay{2010}{}{}.
		\newblock
		\APACrefbtitle {{Ooh! Ahh! Google Images presents a nicer way to surf the
				visual web}.} {{Ooh! Ahh! Google Images presents a nicer way to surf the
				visual web}.}
		\newblock
		\APACaddressPublisher{}{Google Official Blog}.
		\newblock
		\begin{APACrefURL}
			\url{https://googleblog.blogspot.com/2010/07/ooh-ahh-google-images-presents-nicer.html}
		\end{APACrefURL}
		\PrintBackRefs{\CurrentBib}
		
		\bibitem [\protect \citeauthoryear {%
			Stiny%
		}{%
			Stiny%
		}{%
			{\protect \APACyear {2008}}%
		}]{%
			StinyBook}
		\APACinsertmetastar {%
			StinyBook}%
		\begin{APACrefauthors}%
			Stiny, G.%
		\end{APACrefauthors}%
		\unskip\
		\newblock
		\APACrefYear{2008}.
		\newblock
		\APACrefbtitle {{Shape: talking about seeing and doing}} {{Shape: talking about
				seeing and doing}}.
		\newblock
		\APACaddressPublisher{}{The MIT Press}.
		\PrintBackRefs{\CurrentBib}
		
	\end{thebibliography}
		
	%%%%%%%%%%%%%%%%%%%%%%
	
	\newpage
	\appendix
	\section{Implementation  Details}
	\label{sec:tech}
\begin{figure}[h]
		\centering
		\begin{tabular}{c}
			\includegraphics[width=0.95\textwidth]{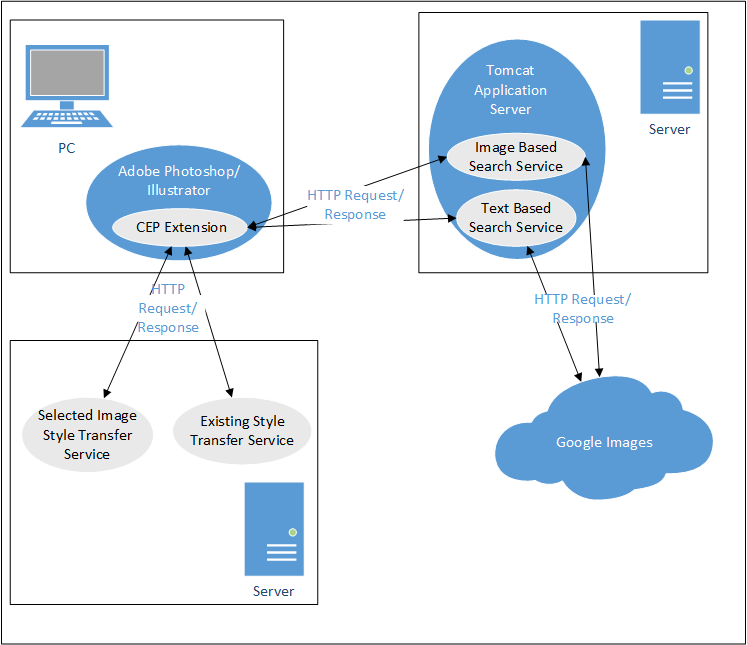}
		\end{tabular}
		\caption{System Architecture}
		\label{fig:tez_2-4}
	\end{figure}

	\noindent 
	The system architecture is depicted in Figure~\ref{fig:tez_2-4}. 
	We used Representational State Transfer (REST)  architectural style for client-server communication. The advantage  is that clients do not need to know the content or prepare a special configuration.
	We used Java's REST libraries for the case of image hunt and Waffle for the case of style adaptation. Web services that satisfy requirements of REST style commonly use stateless HTTP protocol (which is a web protocol) and its standard methods such as GET, POST, PUT, DELETE. All requests sent from the client to the server are POST requests. We have developed web services that take clients' requests as an image or text, pass them to the server, and then return JavaScript Object Notation objects.

	Our custom-built extension to Adobe Creative Cloud that reside in the personal computer of the platform user is developed using the Common Extensibility Platform, a scripting platform that enables developers to extend the functionality of Adobe applications. A Common Extensibility Platform (CEP)  script sends an input expressing the client's request to the server layer and waits for the resulting images (either a set of inspirational images or styles applied via various style transfer processes). 
	There is a respective server layer for each custom-built new application programming interfaces (API) that we have developed for each of the CEP scripts. 
	
	Below, in two subsections, we highlight key implementation challenges  and how we addressed them.
	
	\subsection{Image Hunt}
	\label{app:imagehunt}
	\noindent 
	To perform image hunt tasks, the server layer has to communicate with Google Images (which is on a cloud). 
	Unfortunately, Google does not provide any application programming interface for image-based search. Moreover, the application programming interface for the text-based search is no longer supported. Hence, we had to develop APIs with which we can send an image or keyword to Google images and receive the resulting search results using the plain HTTP protocol. 
	
	This process is a bit tricky. 
	For uploading an image, the server layer API needs to send an HTTP POST request to the Google Reverse Image Search Engine. But the Google Reverse Image Search Engine does not allow/accept any POST requests. Hence, our API sends an HTTP GET request which includes an image link. Of course, the linked images should be stored on a publicly accessible server in order to give permission to the Google Reverse Image Search Engine access to the link. Once the API sends the request with the image link, the Reverse Image Search Engine returns a response containing an HTML context. This is the same result one would get if the Google Images were searched using a regular web browser. That is, the returned context consists of components that are understandable to humans only if they are interpreted by a web browser. Hence, the server layer should have a parsing capability in order to extract image data from the respective components. The parsing process is based on selecting the correct HTML components which contain image links of query results. Each component has an id number and this number is used for determining the component with the image link.  A Java HTML Parser library (JSoup) is used to retrieve the content of the component with the given id. After extracting the resulted image links, our image-based API creates a JavaScript Object Notation array with these links and returns back to the client layer.
	To give more insight, let us examine image-based search in detail.
	Once the platform user clicks refresh button on the user interface at the client layer (our custom-built extension to Adobe), an event is fired for reading the active document and exporting it as PNG image. Adobe's own Common Extensibility Platform readily provides necessary built-in functions to perform these operations. After exporting the current design as a PNG image, custom-built extension reads this image in binary format by using Node.js and transforms it to a binary string to create an appropriate input for the request. The client layer later sends this input to server layer and waits for a response. If the server returns a response of which status is 200, it means that the operation is successful. In that case, the response also includes a JavaScript Object Notation array of resulted image links. The client layer then parses the array and renders these images to the platform user. If the server returns a response with a status different than 200, which means an error occurred, the client layer shows nothing to the platform user. This is the picture from the client-side point of view. Now, let us examine what happens on the server side. Once the client server sends an input in the form of a binary string as explained above, the server layer image-based search application programming interface (a java web application which runs under Apache Tomcat on a remote server) accepts the image.  The image-based search application programming interface acts like a restful service; once the data transmission is complete, it gets ready for querying the Google Reverse Search Engine. Recall that Google Reverse Search Engine does not provide any interface for this search.  Hence, our interface sends a request to Google Reverse Search Engine and parse the returned response as detailed before.

	\subsection{Style Adaptation}
	\label{app:imagestyle}
	\noindent 
	For style adaptation, we employ Neural Style Transfer, a publicly available neural network code  \cite{Johnson2015}.  Unfortunately, there is no web interface for Neural Style Transfer that can respond to client requests. Hence, client-server communication should be again supplied by our system. We use Waffle (\cite{Waffle}, a communication framework for Torch applications. Waffle can listen to POST requests and return JavaScript Object Notation (JSON) objects. Currently, there are two APIs that we have developed using Waffle: Selected Image Style API and Existing Style API. The first one takes two images (style image and content image) as input for transferring the style of the style image to the content image. The second API takes only the content image as input and uses one of the pre-coded styles. It is for applying known artistic styles to the content image. 
	
	The client (via our Adobe Common Extensibility Platform scripts) sends POST requests that contain images coded via JavaScript Object Notation. Server layer APIs listen to POST requests and return the results to the client.
	
	%%%%%%%%
	
	\section{Links to visuals}
	\label{sec:appendix_visuals}
	\subsection*{A Clockwork Orange - I}
	\begin{itemize}
		\item \url{https://thenounproject.com/term/bowler-hat/145955/}
		\item \url{https://datamarcos.blogspot.com/2016/01/blog-agromarcos-consuma-leite-em.html}
		\item \url{http://bop-protocol.org/ontario/how-to-get-limes-out-of-beer-bottles.php}
		\item \url{https://newvitruvian.com/explore/glass-clipart-transparent-background/#gal_post_57_glass-clipart-transparent-background-13.jpg}
		\item \url{https://www.furniturerow.com/fr/Oak-Express/Colors-Dining-Table/prod2220107/}
		\item \url{https://j-toy.ru/wa-data/public/shop/products/50/37/3750/images/13505/a_clockwork_orange_rah_action_figure__alex_delarge__cm-6.280x0.jpg}
		\item \url{https://www.brandsoftheworld.com/logo/clockwork-orange}
		\item \url{https://www.ivoox.com/dia-3-banda-sonora-audios-mp3_rf_23530039_1.html}
	\end{itemize}
	\subsection*{A Clockwork Orange - II}
	\begin{itemize}
		\item \url{http://mariafresa.net/single/2331123.html}
		\item \url{https://www.kisspng.com/png-partners-in-health-health-care-health-system-medic-1069801/}
		\item \url{https://www.tasmeemme.com/en/store-items/people-together-\%7Bwith-hand-up/?item=1035208970}
		\item \url{https://www.gograph.com/photo/teamwork-gg69426752.html}
		\item \url{https://tr.pinterest.com/pin/388998486549179153/}
		\item \url{https://gamesageddon.com/stock/media?id=99786289}
		\item \url{https://foryourimages.com/items/search/smile\%20seeing?page=70}
		\item \url{https://tr.pinterest.com/pin/797840890210087783/}
		\item \url{https://designtwerks.wordpress.com/2013/11/01/adobe-illustrator-tutorial-create-a-death-goddess-inspired-by-mexicos-day-of-the-dead/}
		\item \url{https://tr.pinterest.com/pin/290693350919712167/}  
		\item \url{https://twitter.com/rick\_nmortyc137/status/993572579167293449}
		\item \url{https://www.threadless.com/product/6174/Visit\_Mordor/style,design}
	\end{itemize}

	\subsection*{Desert Monster}
	\begin{itemize}
		\item \url{https://iccup.com/community/thread/1349092.html}
		\item \url{https://www.forient.ch/category/news}
		\item \url{http://blackbackgroundhd.blogspot.com/2013/08/black-emo-background-free-download.html}
		\item \url{https://www.shutterstock.com/search/village+worker}
		\item \url{https://www.tutsela.com/wp-content/uploads/2018/04/?DA}
		\item \url{https://www.deviantart.com/urielperez/art/wallpaper-pulpo-87853650}
	\end{itemize}
	\subsection*{Butterfly}
	\begin{itemize}
		\item \url{http://www.mulierchile.com/butterfly.html/9}
		\item \url{https://www.flickr.com/photos/icemanphotos/9569650958/in/\%7Dgallery-28521388@N06-72157634519011271/}
		\item \url{https://www.needpix.com/photo/143217/daisy-flower-meadow-flowers-nature-plant-summer-meadow-garden-bloom}
		\item \url{https://za.pinterest.com/pin/543106036283664867}
		\item \url{https://wallpaperaccess.com/ocean}
		\item \url{http://backgrounds4k.net/dandelion/}
		\item \url{http://markinternational.info/hawaii-background-wallpaper/222797280.html}
		\item \url{http://markinternational.info/freewallpaper/224060209.html}
	\end{itemize}

	%
	%%%%%

\end{document}